\documentclass[letterpaper,12pt]{article}
\usepackage{tabularx} 
\usepackage{amsmath} 
\usepackage{graphicx}
\usepackage[font=scriptsize]{caption}
\usepackage{subcaption}
\usepackage[margin=1in,letterpaper]{geometry} 
\usepackage{cite} 
\usepackage[final]{hyperref} 
\hypersetup{
	colorlinks=true,       
	linkcolor=blue,        
	citecolor=blue,        
	filecolor=magenta,     
	urlcolor=blue         
}
\usepackage{blindtext}
\allowdisplaybreaks

\begin{document}

\title{Analyzing the Effects of Fifth and Seventh Order Terms in a Generalized Henon-Heiles Potential}
\author{Nandana Madhukara}
\date{}
\maketitle

\begin{abstract}
In 1962, astronomers Michel Hénon and Carl Heiles studied orbits of stars around centers of galaxies to determine the third integral of motion in galactic dynamics. In order to do this, they reduced the system down to a 2-dimensional axisymmetric Hamiltonian system. Now this is known as the Hénon-Heiles (HH) System. Due to its apparent simplicity but extremely complicated dynamical behavior, this system is currently a paradigm in dynamical systems.

In this paper, we perform a series expansion up to the seventh order of a potential with axial and reflection symmetries. After some transformations, this turns into the generalized Hénon-Heiles (GHH) system where we separate the fifth and seventh-order terms. We qualitatively analyze this system for energies near the threshold between bounded and unbounded motion with Poincaré sections and quantitatively analyze with Lyapunov Exponents. We find that particles far from the critical energy demonstrate less chaos. Additionally, the fifth-order term creates more regularity while the seventh-order term does the opposite.
\end{abstract}

\section{Introduction}

There has always been an interest in the existence of a third integral of galactic motion or a quantity that is constant during the motion like energy or angular momentum. Specifically, suppose we have a time-independent and axisymmetric gravitational potential of a galaxy. If we consider it in cylindrical coordinates, the motion will be in the 6-dimensional phase space $(r, \theta, z, \dot{r}, \dot{\theta}, \dot{z})$. This means that there should be five integrals of motion, $I_j$, such that \[I_j(r, \theta, z, \dot{r}, \dot{\theta}, \dot{z}) = C_j\] where $j = 1$ to $5$ and the $C_j$s are constants. Of these five integrals, $I_4$ and $I_5$ are non-isolating and have no significance so they are disregarded. Now $I_1$ and $I_2$ are the well-known energy and angular momentum so the question is: what is the nature of $I_3$?

In 1962, to solve this problem, astronomers Michel Hénon and Carl Heiles \cite{Henon_Heiles_1964} came up with a system described by the following time-independent Hamiltonian: \[\mathcal{H} = \frac{1}{2}(\dot{x}^2 + \dot{y}^2) + \frac{1}{2}(x^2+y^2) + x^2y - \frac{y^3}{3}.\] The beauty of this system is that it is analytically very simple but produces very interesting and complicated trajectories. This has caused the original paper to be one of the most cited works in nonlinear dynamics and now this system's applications range from the study of geodesics in general relativity \cite{Dolan_Shipley_2016} to quantum entanglement using squeezed coherent states \cite{Joseph_Sanjuán_2016}.

Lots of work has been done generalizing the system. Around 1980, Frank Verhulst \cite{verhulst1979discrete} performed a series expansion up to the fourth order of a general potential with axial and reflection symmetry now known as the Verhulst Potential. The orbital structure in the center of a triaxial galaxy with an analytical core \cite{deZeeuw_1985}, the correlation between the Lyapunov exponents and the size of the chaotic areas in the surface of section \cite{Chandler:1989:BQS}, or the escape zones in a quartic potential \cite{Barbanis_1990} have all been studied using the Verhulst potential. Recently, the potential has been expanded to fifth order \cite{Dubeibe_Riaño-Doncel_Zotos_2018} and seventh order \cite{Dubeibe_Zotos_Chen_2020}.

In this paper, we perform a series expansion of an axisymmetric potential up to the seventh order resulting in the GHH-system. In order to study the transition from the HH to the GHH system, we introduce parameters for higher-order terms. Unlike the single parameter used in \cite{Dubeibe_Zotos_Chen_2020}, we introduce two parameters, $\delta$ and $\alpha$, for the fifth and seventh order terms, respectively, as presented in \cite{Zotos_Dubeibe_Riaño-Doncel_2021}. Additionally, we will study systems with energies near but below the threshold between bounded and unbounded motion. 

The paper is organized as follows: In Section \ref{Background}, we will give some background on the type of systems we will be studying and the tools we will be using. Next in Section \ref{Generalized} we will derive the GHH system and in Section \ref{Results} we will calculate the critical energy levels between bounded and unbounded motion. Then we will look at Poincaré sections and Lyapunov Exponents of systems with values for $\delta$, $\alpha$, and the energy. We will end the paper with conclusions we draw from the results.

\section{Background}\label{Background}

\subsection{Bounded and Unbounded Orbits}

\begin{figure}[h]
\centering
\includegraphics[width=0.35\textwidth]{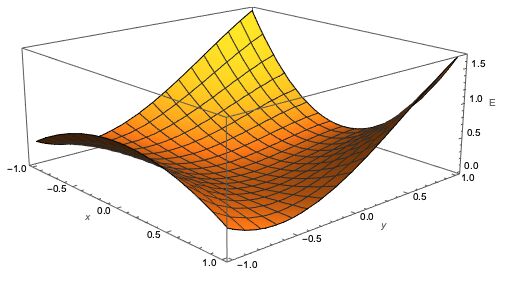}
\caption{HH-System Potential Landscape}
\label{fig:equipotential}
\end{figure}

In this project, we study bounded and unbounded orbits of particles in the GHH System. To define bounded, we first plot the potential energy as a function of $x$ and $y$. For example, Figure \ref{fig:equipotential} is the plot of the HH potential.

We call this surface a potential landscape. Now we can use this landscape to study the dynamics of objects under the influence of the potential. The main thing to note is that there can be valleys or local minimums in the surface where the object can get trapped. Now we are reading to define what bounded motion is.

\begin{figure}[h]
     \begin{center}
     \begin{subfigure}[b]{0.35\textwidth}
         \centering
         \includegraphics[width=\textwidth]{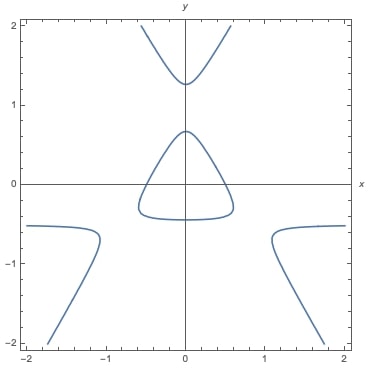}
         \caption{}
         \label{fig:E=1/8}
     \end{subfigure}
     \hspace{0.1\textwidth}
     \begin{subfigure}[b]{0.35\textwidth}
         \centering
         \includegraphics[width=\textwidth]{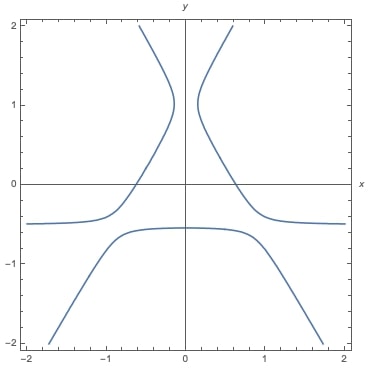}
         \caption{}
         \label{fig:E=1/5}
     \end{subfigure}
    \end{center}
    \caption{Contour plot at (a) $E=1/8$ and (b) $E=1/5$}
    \label{fig:three graphs}
\end{figure}

Let the object start at an energy $E$. As an example, Figure \ref{fig:E=1/8} would be the contour plot of the HH potential at $E=1/8$. Notice the triangular loop in the center. If the object starts on that loop, it would get trapped it would never escape. So we call the orbital motion bounded at $E=1/8$. However, if $E$ is greater than some critical value, $E=1/5$, the contour plot would be like Figure \ref{fig:E=1/5}.

\subsection{Poincaré Sections}
To analyze the chaos, we use Poincaré sections. The idea of Poincaré sections is to slice a higher dimensional phase space into a lower dimensional space. Phase space is a space that represents all possible states of the system. So, the dynamics of a particle in a system can be thought of as movement in phase space. In our case, the phase space is 4 dimensional since its axes would be $x$, $y$, $\dot{x}$, $\dot{y}$. Therefore, to visualize it more easily, we slice it until we get a 2-dimensional space which is our Poincaré section.

\begin{figure}
     \centering
     \begin{subfigure}[b]{0.35\textwidth}
         \centering
         \includegraphics[width=\textwidth]{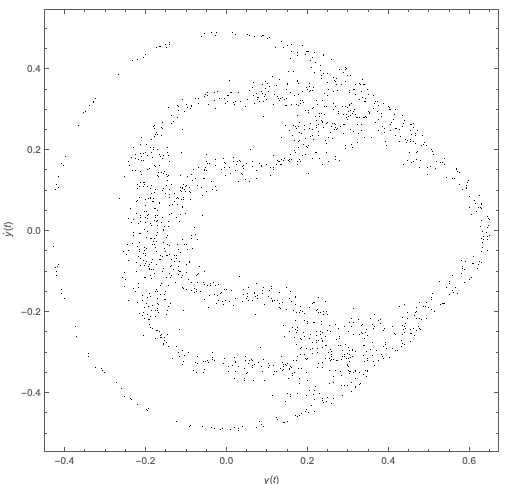}
         \caption{}
         \label{fig:onepy_0}
     \end{subfigure}
     \hspace{0.1\textwidth}
     \begin{subfigure}[b]{0.35\textwidth}
         \centering
         \includegraphics[width=\textwidth]{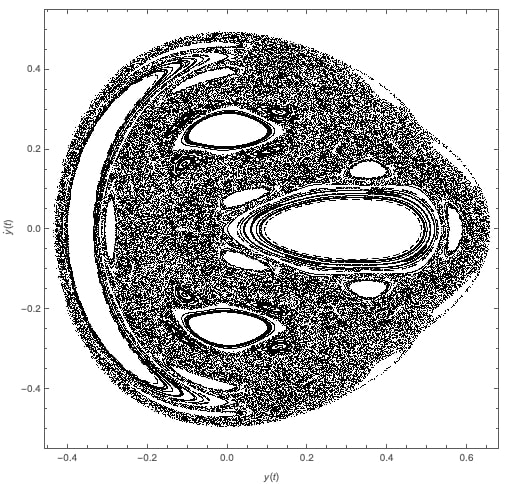}
         \caption{}
         \label{fig:manypy_0}
     \end{subfigure}
    \caption{Poincaré sections of HH-System at $E=1/8$}
\end{figure}

For initial conditions, we set $x_0 = 0.1$, $y_0=0$, we pick a random value for $\dot{y}_0$, and we find $\dot{x}_0$ by using conservation of energy: $H = E$. We first slice our phase space to get rid of $\dot{x}$ giving us a 3-dimensional space. Then we slice again to only consider the points when $x=0$. When we plot $\dot{y}$ vs $y$, we get Figure \ref{fig:onepy_0} which is for the HH system when $E = 1/8$. Each point represents the $y$ position and velocity at each instant in time when $x=0$. Now we overlay the plot for different values of $\dot{y}_0$ so that we account for all allowed motion. This gives us Figure \ref{fig:manypy_0}.

\subsection{Lyapunov Characteristic Exponents}

\begin{figure}[h]
\centering
\includegraphics[width=0.5\textwidth]{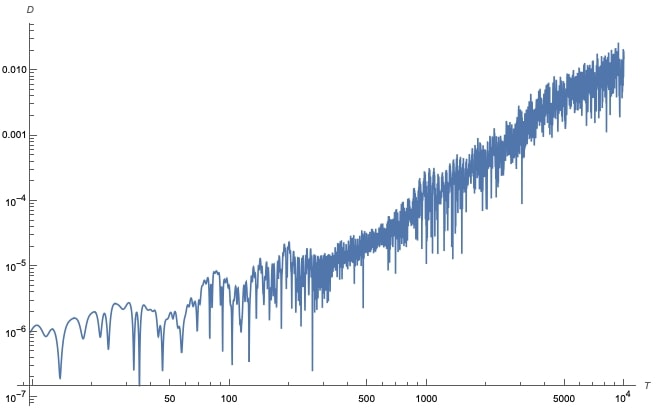}
\caption{Contour Plot of HH Potential at $E=1/8$}
\label{fig:Lyapunov_Distance}
\end{figure}

Another tool we use to analyze the chaos is Lyapunov Characteristic Exponents (LCEs) which are a way of making chaos quantitative. Specifically in this project, we look at maximal LCEs. The main idea behind these exponents is to take two close trajectories and see how they grow apart or converge. If we take the Hénon-Heiles system, we can observe how two extremely close initial points grow apart. For our distance metric, we use the normal Euclidean distance between two points. When $E=1/8$ and the initial separation is $D_0 = 10^{-6}$, we get a log-log plot of the distance as shown in Figure \ref{fig:Lyapunov_Distance}.
If we let the distance $D = D_0e^{\lambda t}$, we can define $\lambda$ to be our maximal LCE. Since we are using a log-log plot, we can find $\lambda$ by finding the slope of the line fit to this graph. In chaotic motion, two close initial points grow apart exponentially so a positive $\lambda$ indicates chaotic motion. The larger $\lambda$ is, the more chaotic the motion is.

\section{Generalized Hénon-Heiles Potential}\label{Generalized}

In this paper, we use the same GHH Hamiltonian as the one in \cite{Zotos_Dubeibe_Riaño-Doncel_2021}. For the sake of completeness, we shall present the derivation. To do so, we first start with the Hamiltonian of a general axisymmetric potential: \[\mathcal{H} = \frac{1}{2}(\dot{r}^2 + \dot{z}^2) + V(r, z) + \frac{L_z^2}{2r^2}\] where $L_z$ is the angular momentum about the $z$-axis. 
Now we define $V_{\text{eff}} = V(r, z) + L_z^2/(2r^2)$ which has a minimum at $(r, z) = (r_0, 0)$. We express $V_{\text{eff}}$ has a seventh order Taylor Series around $(r_0, 0)$ to get 
\begin{align*}
    V_{\text{eff}}(r, z) &\approx a_1\xi^4 + z^4(a_2+b_2\xi+c_3\xi^2+d_4\xi^3)+z^2 \\
    &\times (a_3\xi^2 + b_3\xi^3+c_4\xi^4+d_3\xi^5+\omega_2^2+\xi\epsilon) \\
    &+ \beta\xi^3+b_1\xi^5+c_1\xi^6+z^6(c_2+d_2\xi) \\
    &+ d_1\xi^7+\xi^2\omega_1^2
\end{align*}
where 
\begin{align*}
    &\xi = r-r_0, \\
    &\omega_1^2=\frac{3L_z}{2r_0^4}+\frac{1}{2}\left.\frac{\partial^2 V_{\text{eff}}}{\partial r^2}\right|_{(r_0, 0)}, \omega_2^2 = \frac{1}{2} \left. \frac{\partial^2 V_{\text{eff}}}{\partial z^2} \right|_{(r_0, 0)}, \\
    &\epsilon = -\frac{1}{2} \left. \frac{\partial^3 V_{\text{eff}}}{\partial r \partial z^2} \right|_{(r_0, 0)}, \beta = -\frac{2L_z^2}{r_0^5} + \frac{1}{6} \left. \frac{\partial^3 V_{\text{eff}}}{\partial r^3} \right|_{(r_0, 0)}, \\
    &a_1 = \frac{5L_z^2}{2r_0^6}+\frac{1}{24} \left. \frac{\partial^4 V_{\text{eff}}}{\partial r^4} \right|_{(r_0, 0)}, a_2 = \frac{1}{24} \left. \frac{\partial^4 V_{\text{eff}}}{\partial z^4} \right|_{(r_0, 0)}, a_3=\frac{1}{4} \left. \frac{\partial^4 V_{\text{eff}}}{\partial r^2 \partial z^2} \right|_{(r_0, 0)}, \\
    &b_1 = -\frac{3L_z^2}{r_0^7}+\frac{1}{120} \left. \frac{\partial^5 V_{\text{eff}}}{\partial r^5} \right|_{(r_0, 0)}, b_2 = \frac{1}{24} \left. \frac{\partial^5 V_{\text{eff}}}{\partial r \partial z^4} \right|_{(r_0, 0)}, b_r = \frac{1}{12} \left. \frac{\partial^5}{\partial r^3 \partial z^2} \right|_{(r_0, 0)}, \\
    &c_1 = \frac{7L_z^2}{12r_0^8} + \frac{1}{720} \left. \frac{\partial^6 V_{\text{eff}}}{\partial r^6} \right|_{(r_0, 0)},  c_2 = \frac{1}{720} \left. \frac{\partial^6 V_{\text{eff}}}{\partial z^6} \right|_{(r_0, 0)}, \\
    &c_3 = \frac{1}{48} \left. \frac{\partial^6 V_{\text{eff}}}{\partial r^2 \partial z^4} \right|_{(r_0, 0)}, c_4 = \frac{1}{48} \left. \frac{\partial^6 V_{\text{eff}}}{\partial r^4 \partial z^2}\right|_{(r_0, 0)}\\
    &d_1 = -\frac{4L_z^2}{r_0^9} + \frac{1}{5040} \left. \frac{\partial^7 V_{\text{eff}}}{\partial r^7} \right|_{(r_0, 0)}, d_2 = \frac{1}{720} \left. \frac{\partial^7 V_{\text{eff}}}{\partial r \partial z^6} \right|_{(r_0, 0)}, \\
    &d_3 = \frac{1}{240} \left. \frac{\partial^7 V_{\text{eff}}}{\partial r^5 \partial z^2} \right|_{(r_0, 0)}, d_4 = \frac{1}{144} \left. \frac{\partial^7 V_{\text{eff}}}{\partial r^3 \partial z^4} \right|_{(r_0, 0)}.
\end{align*}

To better resemble the original Hénon-Heiles system, we introduce the parameters $\alpha$ and $\delta$ and make some replacements: $z \to x$, $\xi \to y$, $a_1=a_2=b_1=-b_2=-b_3=-\delta$, $a_3=-2\delta$, $c_1=c_2=d_1=d_2=d_3=d_4=2\alpha$, $c_3=c_4=\alpha$, $\omega_1=\omega_2 = 1/\sqrt{2}$, $\beta = -1/3$ and $\epsilon = 1$. This gives us the GHH Hamiltonian:
\begin{align*}
\mathcal{H} &= \frac{1}{2}(\dot{x}^2 + \dot{y}^2) + \frac{1}{2}(x^2 + y^2) + x^2y - \frac{y^3}{3} \\
&+ \delta[x^4(y-1) + x^2(y-2)y^2-y^4(y-1)] \\
&+ \alpha[2x^6(y+1) + x^4y^2(2y+1) + x^2y^4(2y+1) + 2y^6(y+1)].
\end{align*}


\section{Results}\label{Results}
\begin{figure}[h]
     \centering
     \begin{subfigure}[b]{0.48\textwidth}
         \centering
         \includegraphics[width=\textwidth]{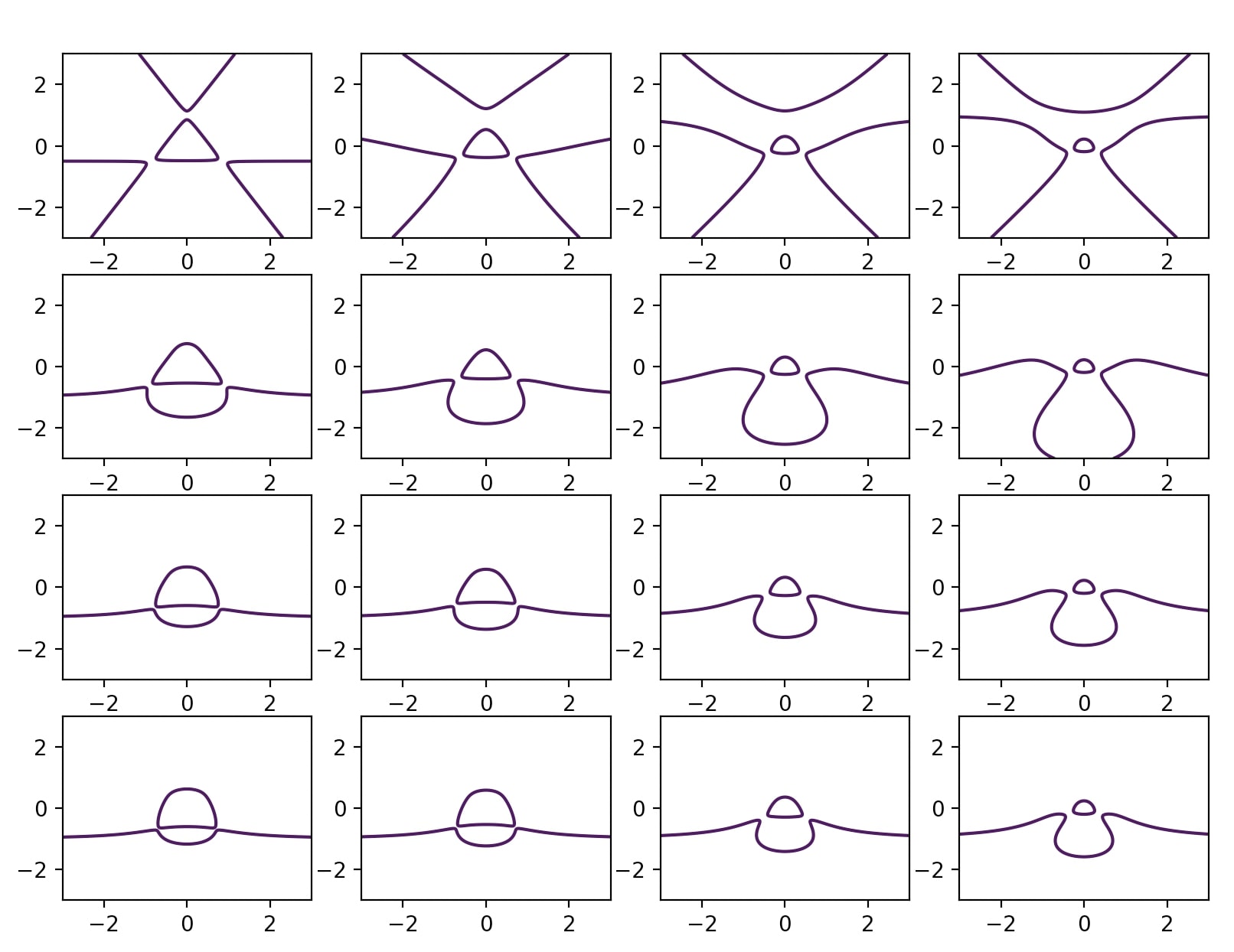}
     \end{subfigure}
     \hfill
     \begin{subfigure}[b]{0.48\textwidth}
         \centering
         \includegraphics[width=\textwidth]{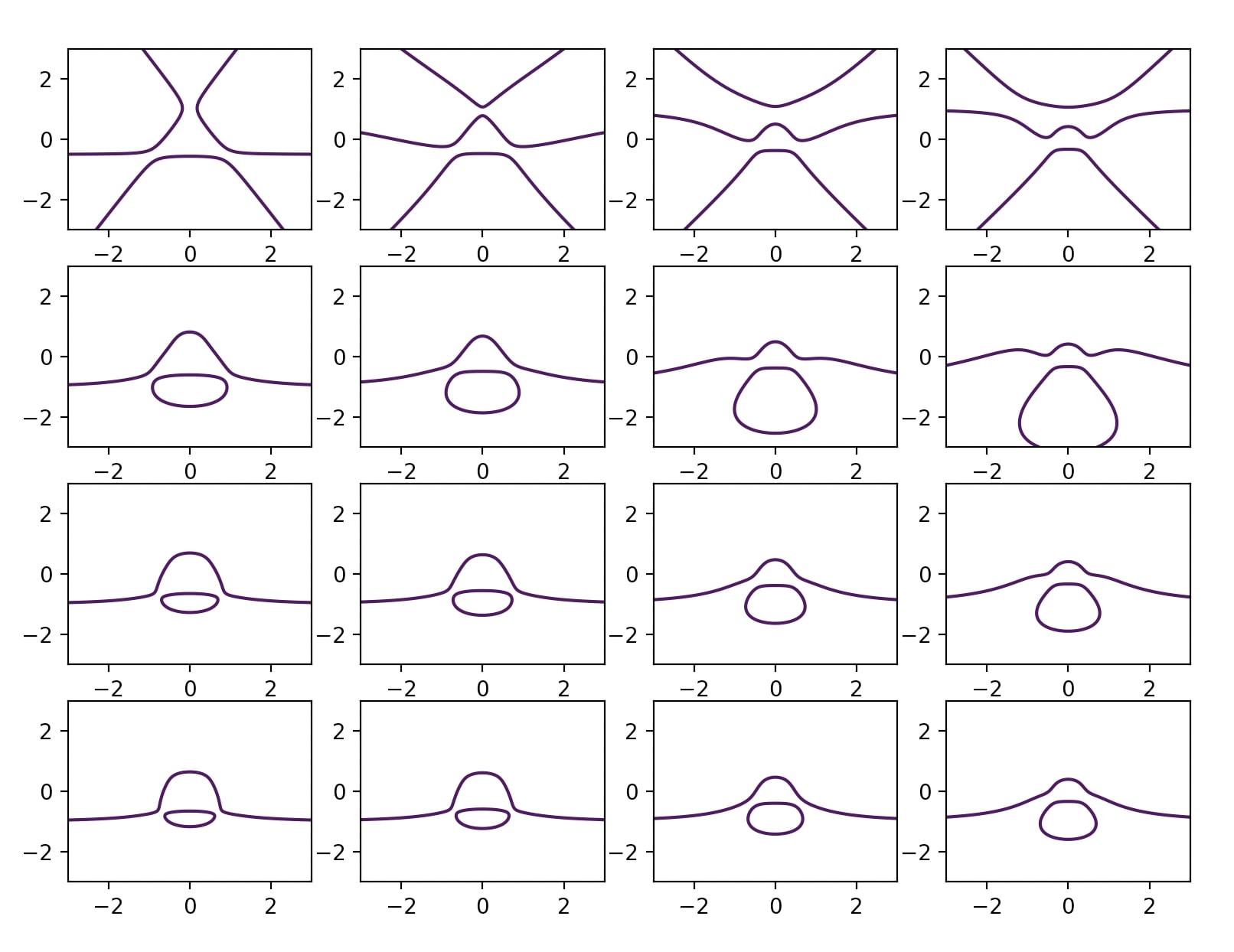}
     \end{subfigure}
    \caption{Transition from Bounded to Unbounded Motion}
    \label{fig:bounded_unbounded}
\end{figure}

The first step we take is to find the boundary between bounded and unbounded motion (when we start near the origin). This depends on $\alpha$ and $\delta$ so we get the following table of energies at the boundary: 
\begin{center}
\begin{tabular}{c| c c c c} 
 $\alpha \text{ } \symbol{92} \text{ } \delta$ & 0 & 0.1 & 0.5 & 1 \\
 \hline
 0 & 0.1666 & 0.1046 & 0.0501 & 0.0323\\ 
 0.1 & 0.2103 & 0.1168 & 0.0513 & 0.0326\\
 0.5 & 0.2756 & 0.1843 & 0.0571 & 0.0339\\
 1 & 0.3086 & 0.242 & 0.0693 &  0.0359
\end{tabular}
\end{center}
We call these energies $E_{\text{min}}$ and Figure \ref{fig:bounded_unbounded} shows the contour plots when the energy is less than and greater than $E_{\text{min}}$.

\begin{figure}[htbp]
    \centering
    \subfloat[]{
        \label{fig:n=1,d=0}
        \includegraphics[width=0.3\textwidth]{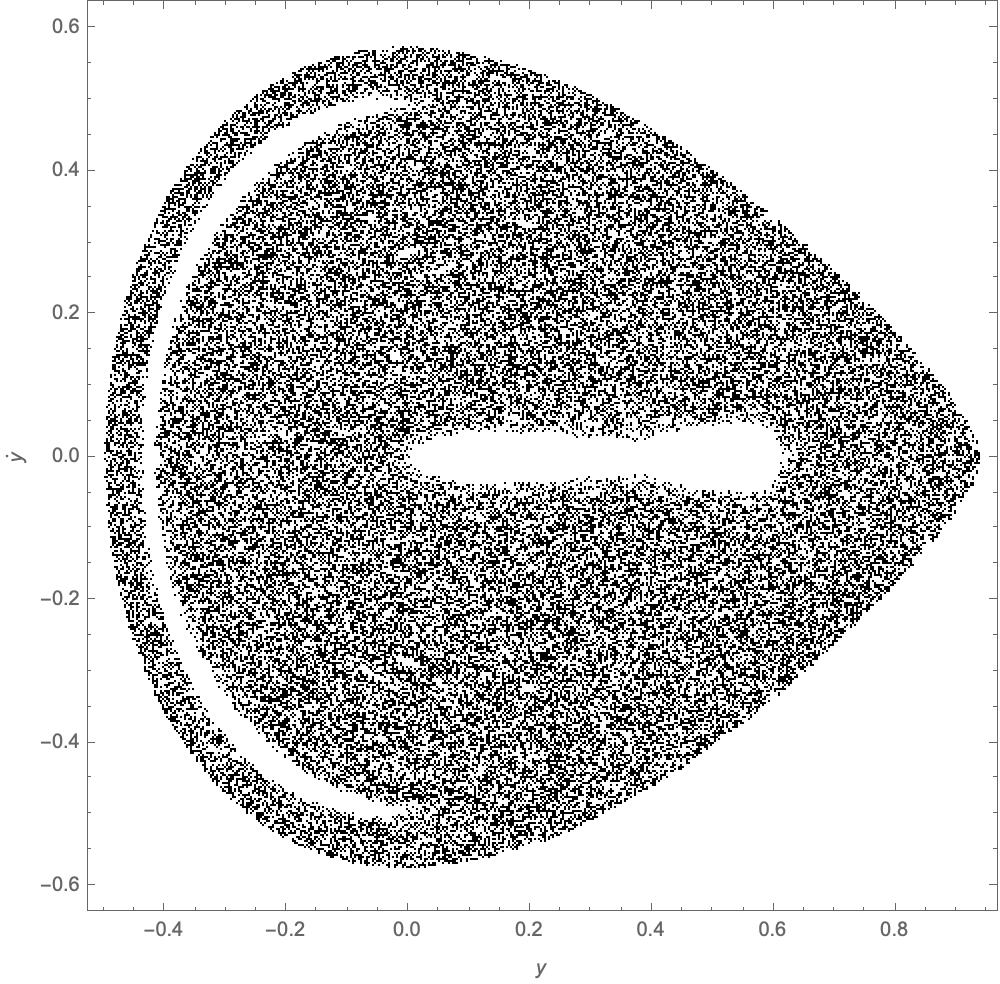}
    }
    \subfloat[]{
        \includegraphics[width=0.3\textwidth]{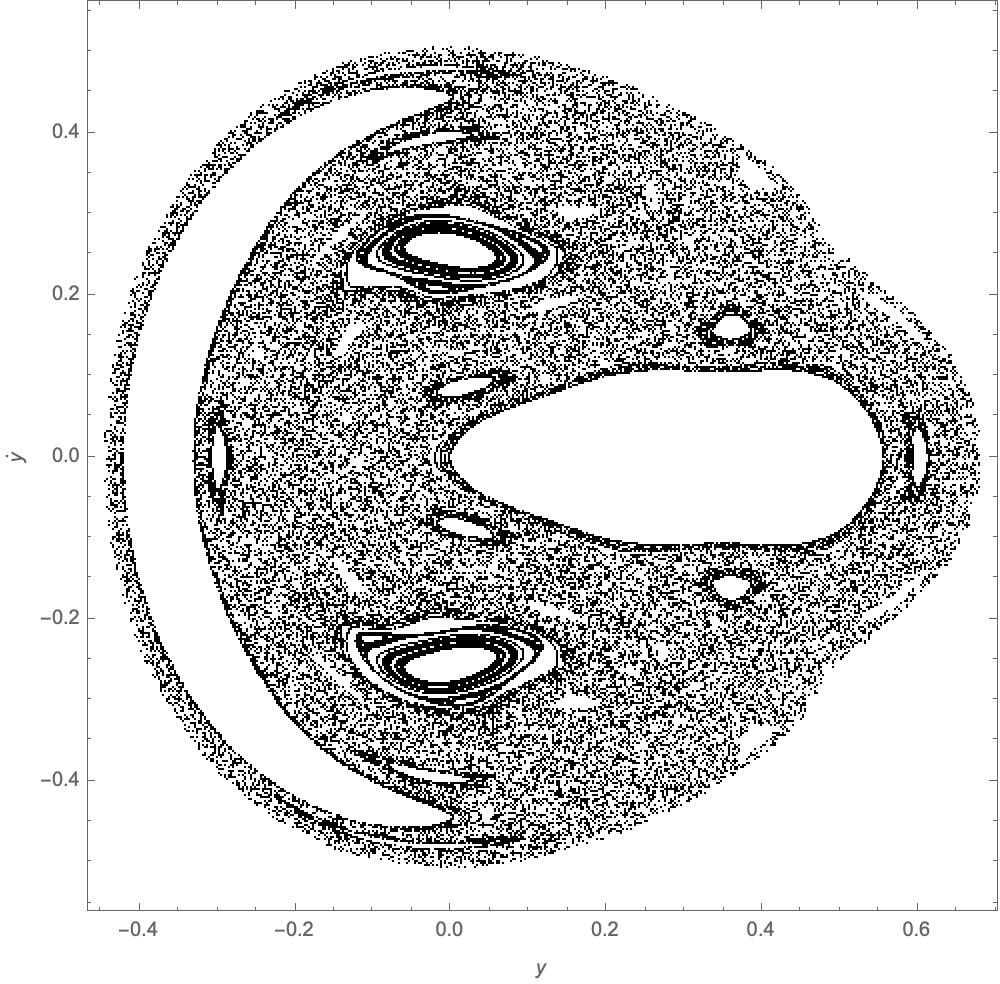}
    }
    \subfloat[]{
        \label{fig:n=41,d=0}
        \includegraphics[width=0.3\textwidth]{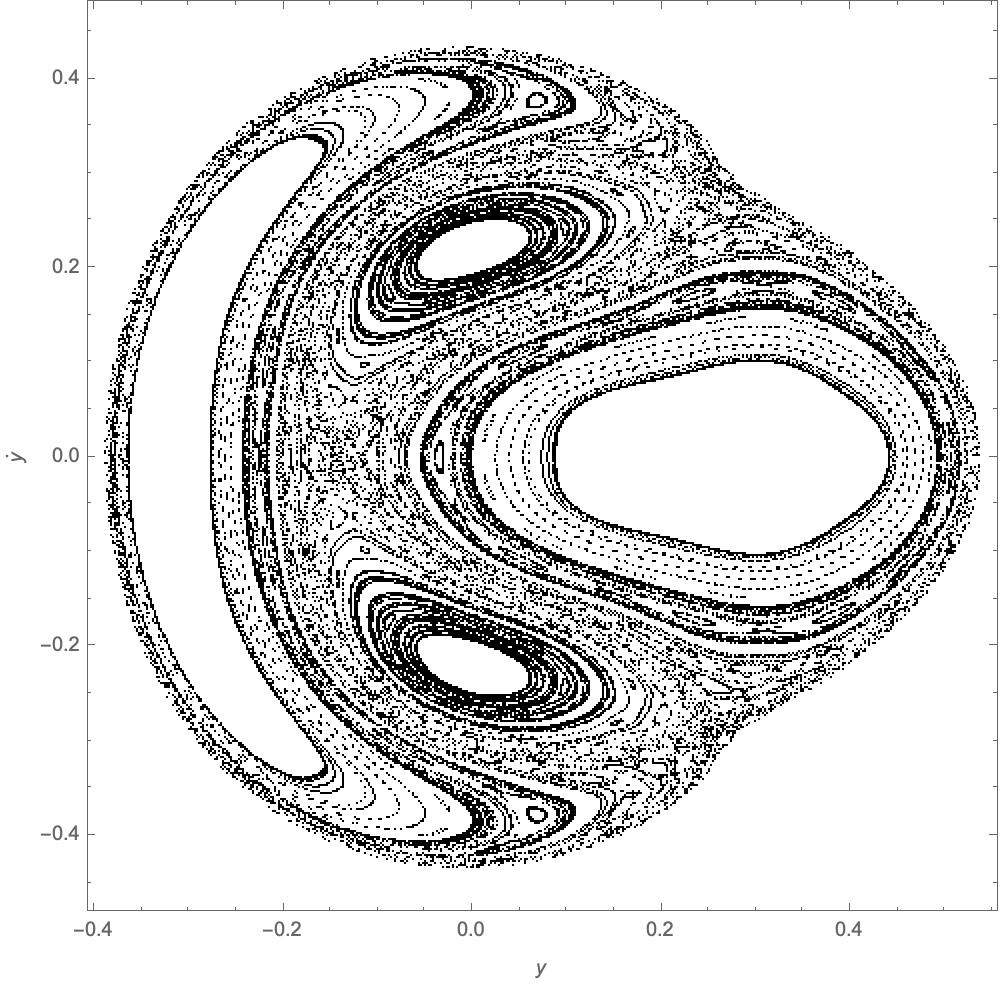}
    }
    \hfill
    \subfloat[]{
        \includegraphics[width=0.3\textwidth]{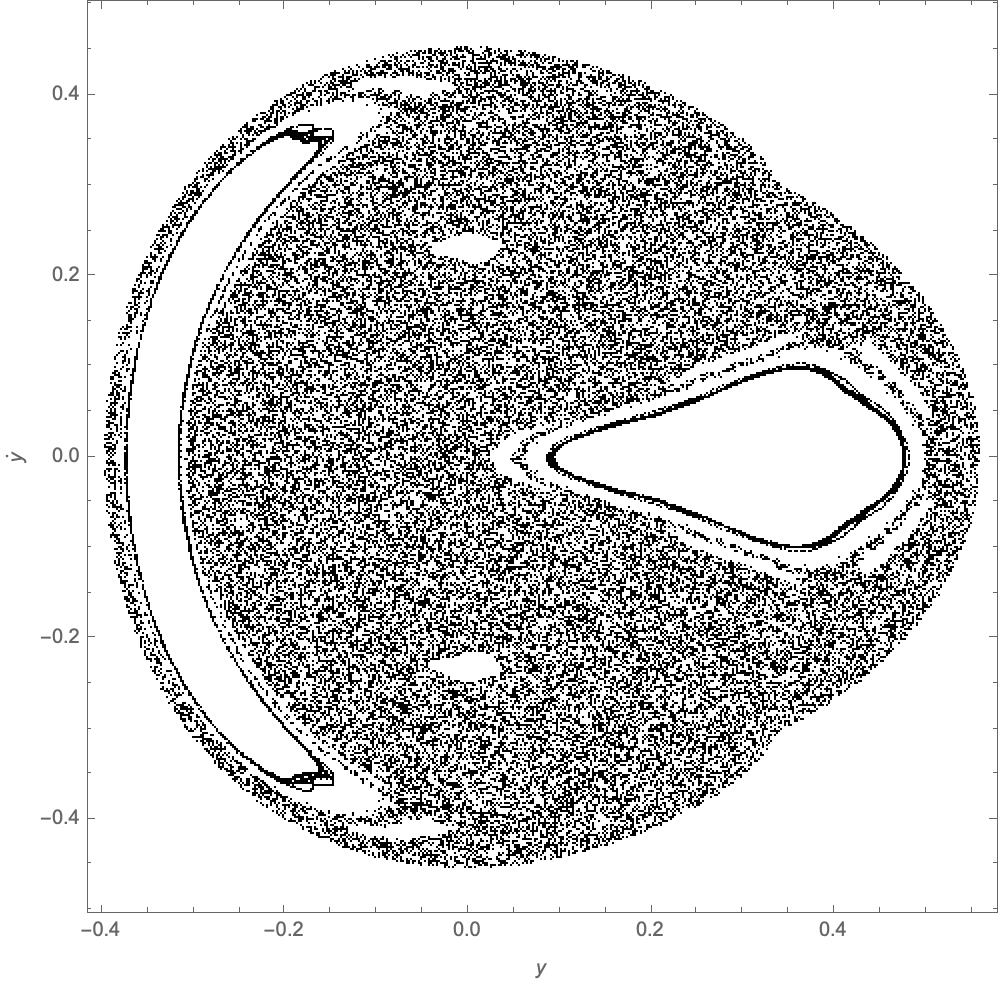}
    }
    \subfloat[]{
        \includegraphics[width=0.3\textwidth]{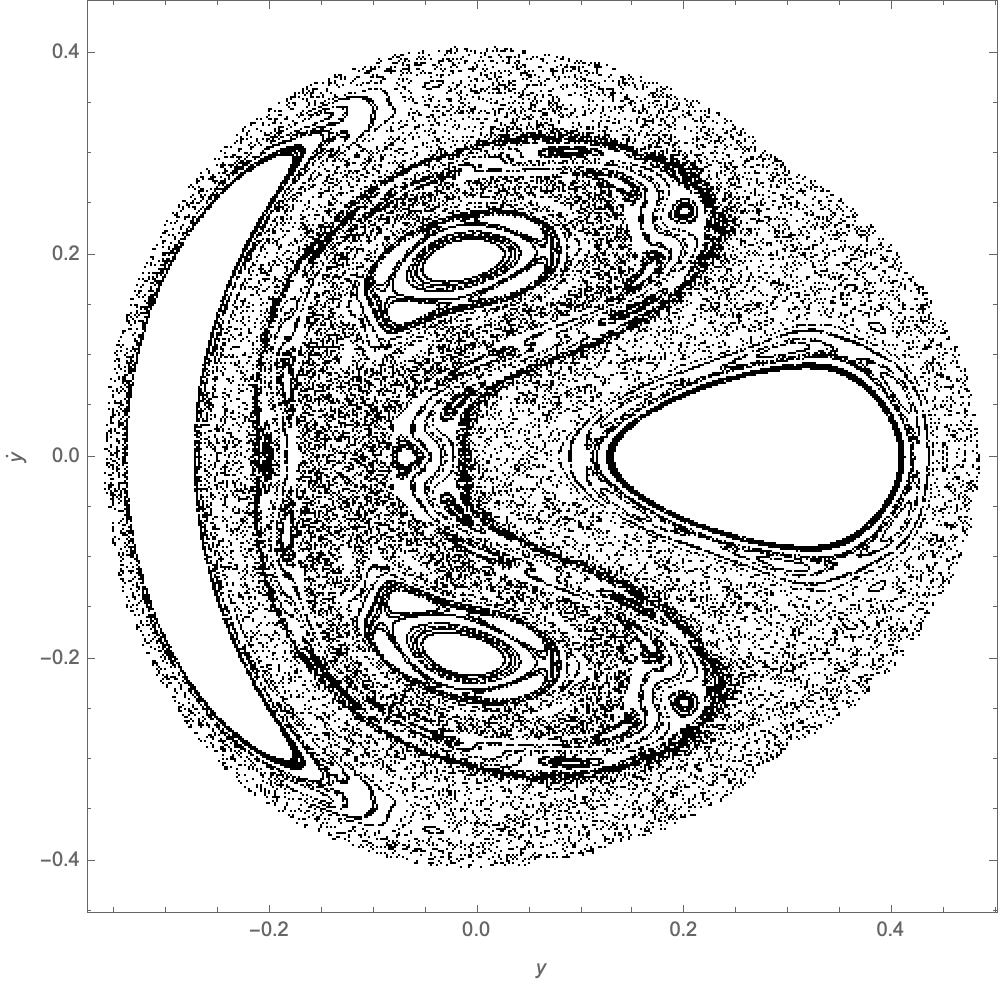}
    }
    \subfloat[]{
        \includegraphics[width=0.3\textwidth]{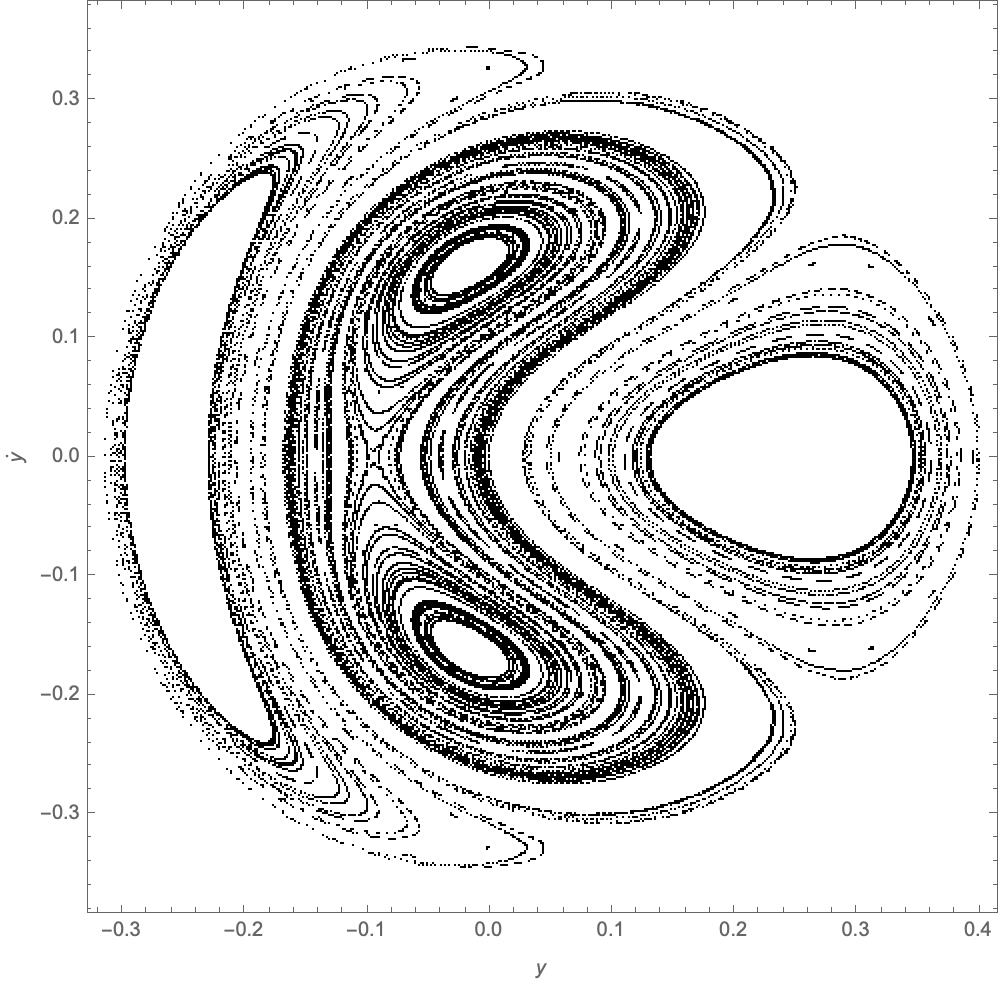}
    }
    \hfill
    \subfloat[]{
        \includegraphics[width=0.3\textwidth]{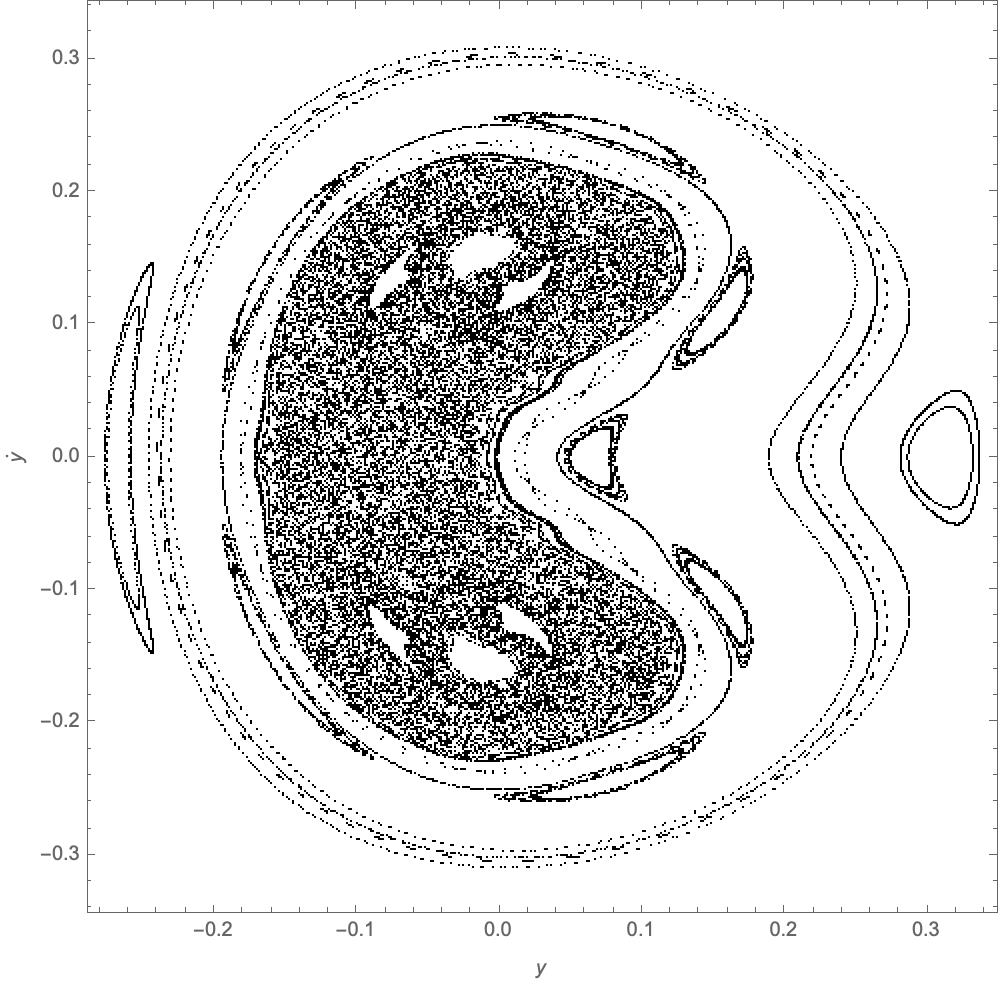}
    }
    \subfloat[]{
        \includegraphics[width=0.3\textwidth]{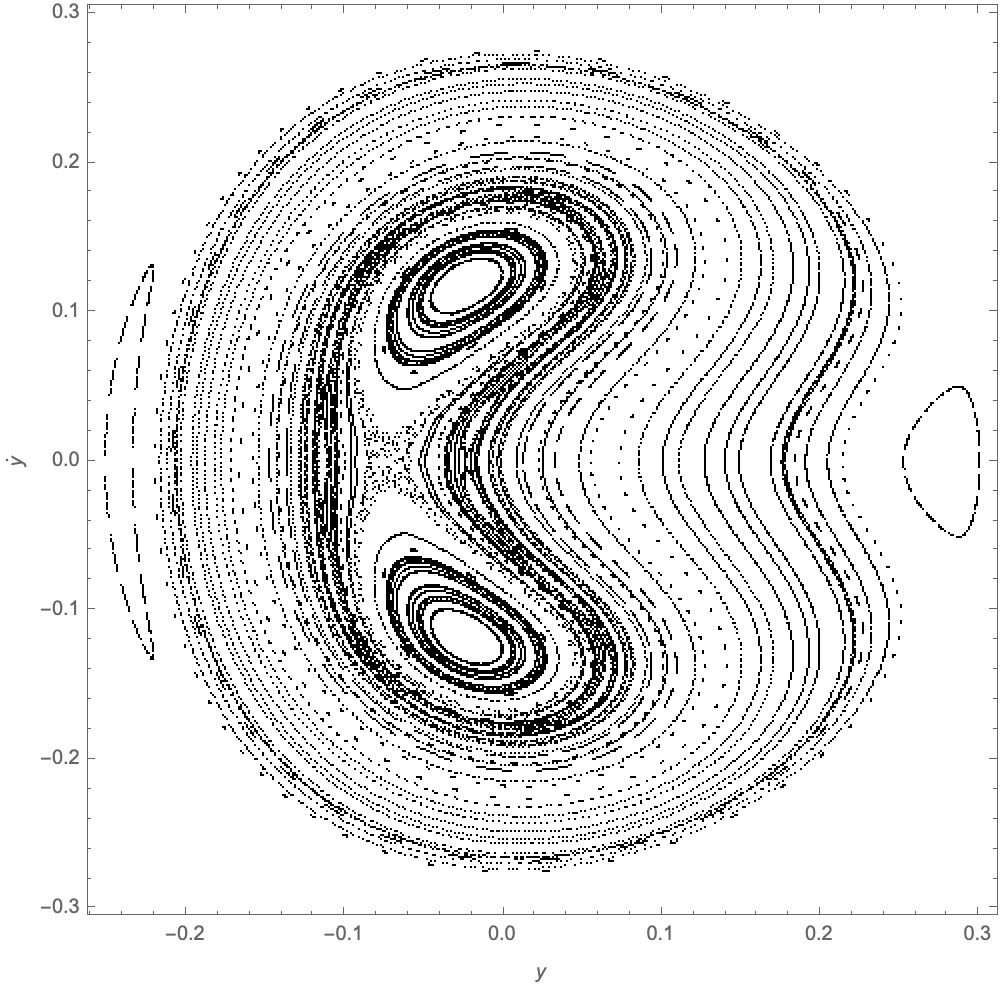}
    }
    \subfloat[]{
        \includegraphics[width=0.3\textwidth]{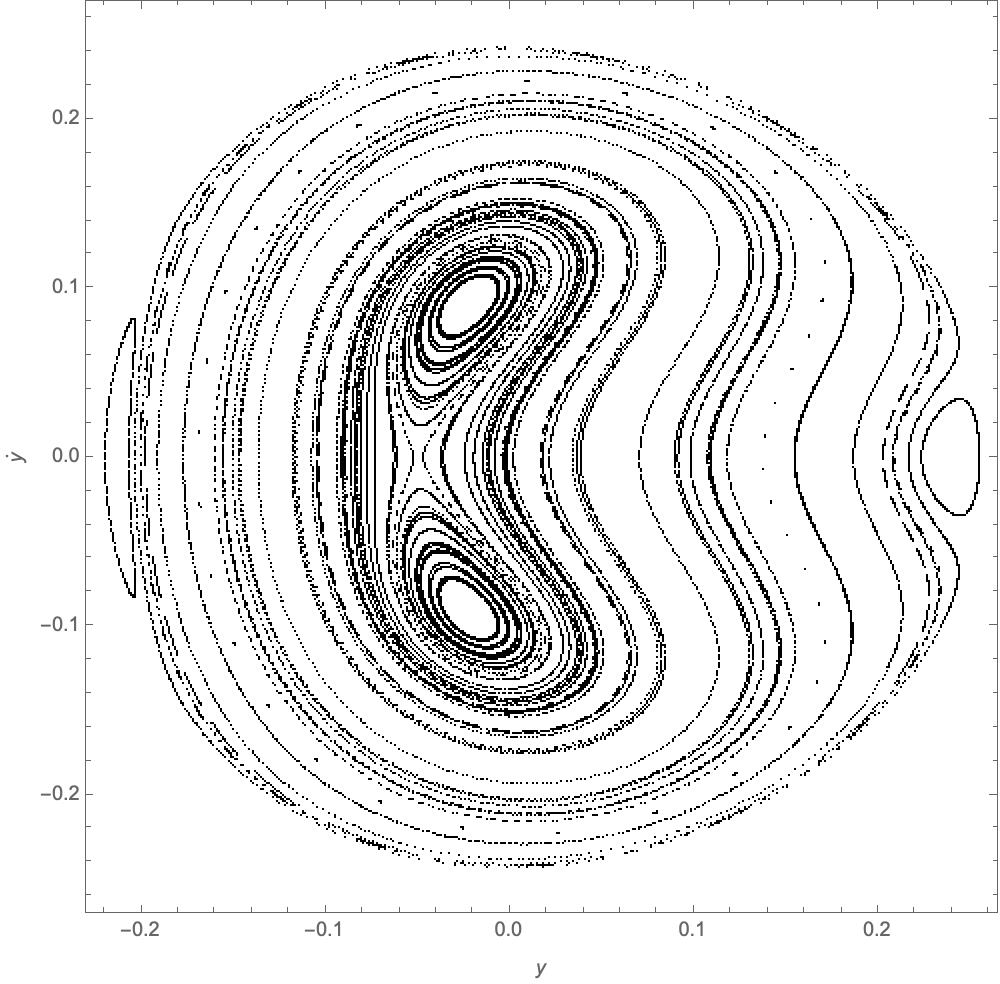}
    }
    \hfill
    \subfloat[]{
        \includegraphics[width=0.3\textwidth]{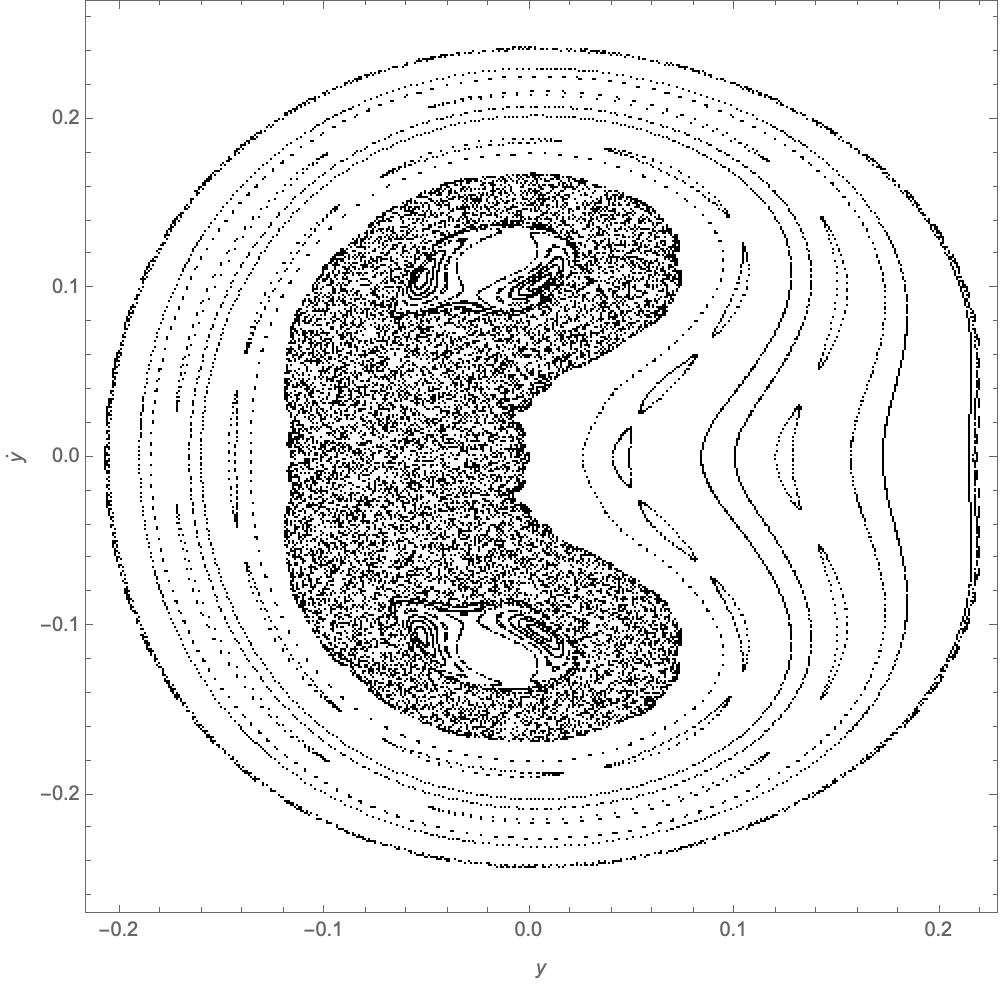}
    }
    \subfloat[]{
        \includegraphics[width=0.3\textwidth]{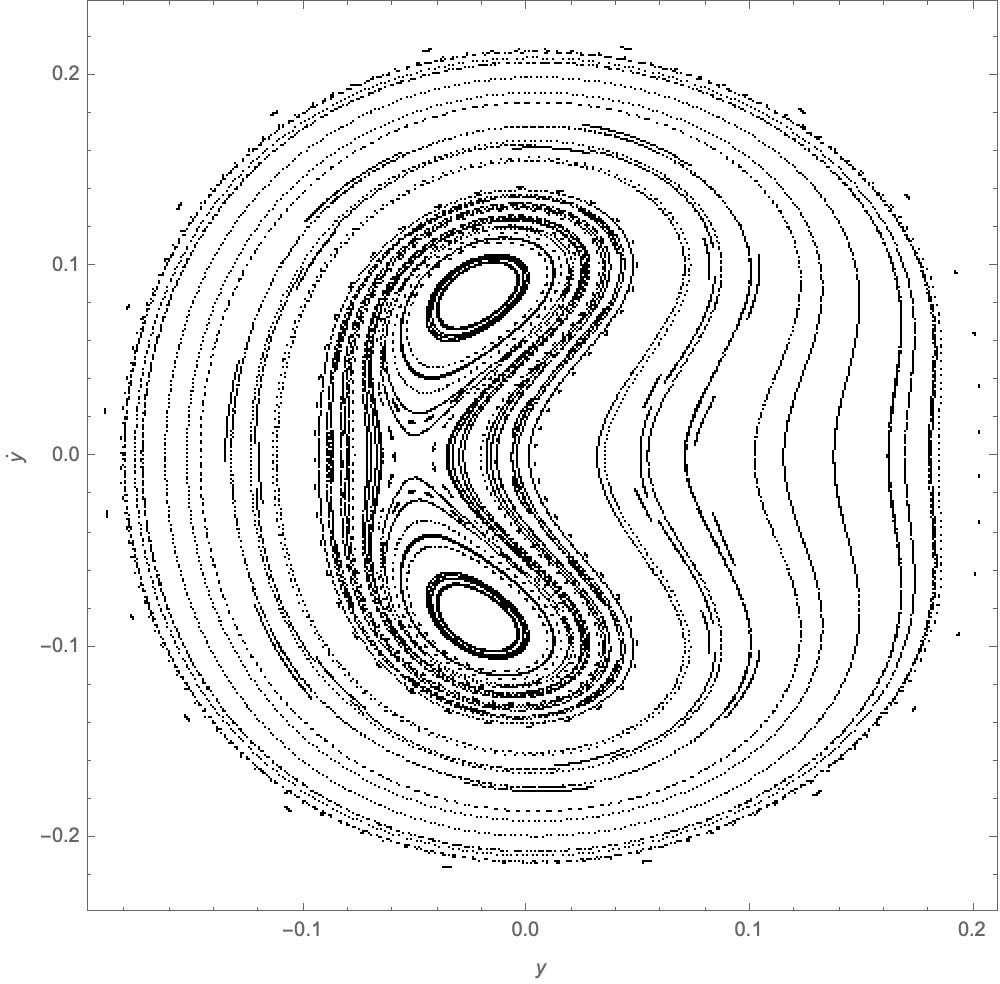}
    }
    \subfloat[   ]{
        \includegraphics[width=0.3\textwidth]{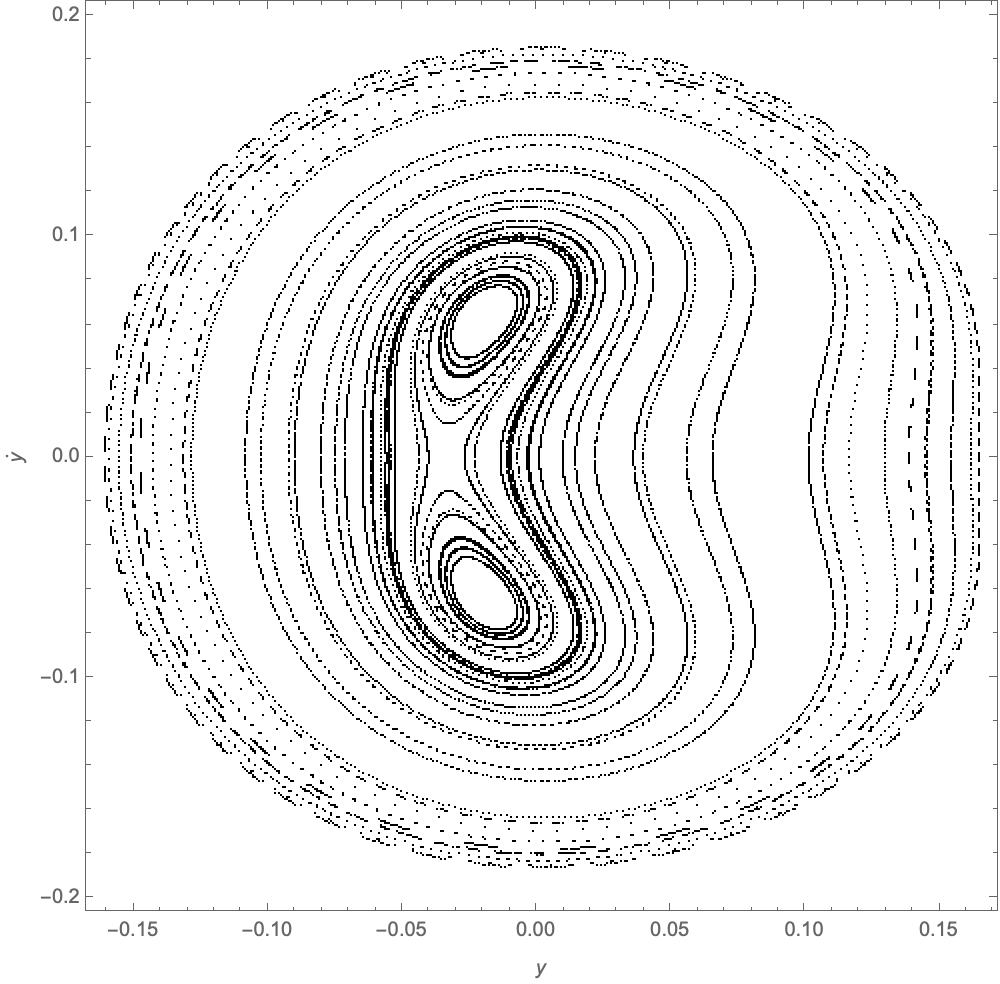}
    }
    \caption{Poincaré sections when $\alpha = 0$ and (a) $\delta = 0$ and $n = 1$, (b) $\delta = 0$ and $n = 21$, (c) $\delta = 0$ and $n = 41$, (d) $\delta = 0.1$ and $n = 1$, (e) $\delta = 0.1$ and $n = 21$, (f) $\delta = 0.1$ and $n = 41$, (g) $\delta = 0.5$ and $n = 1$, (h) $\delta = 0.5$ and $n = 21$, (i) $\delta = 0.5$ and $n = 41$, (j) $\delta = 1.0$ and $n = 1$, (k) $\delta = 1.0$ and $n = 21$, (l) $\delta = 1.0$ and $n = 41$ where the energy for each panel is $E = E_{\text{min}}(1-n/100)$.}
    \label{fig:alpha=0}
\end{figure}

\begin{figure}[htbp]
    \centering
    \subfloat[]{
        \includegraphics[width=0.3\textwidth]{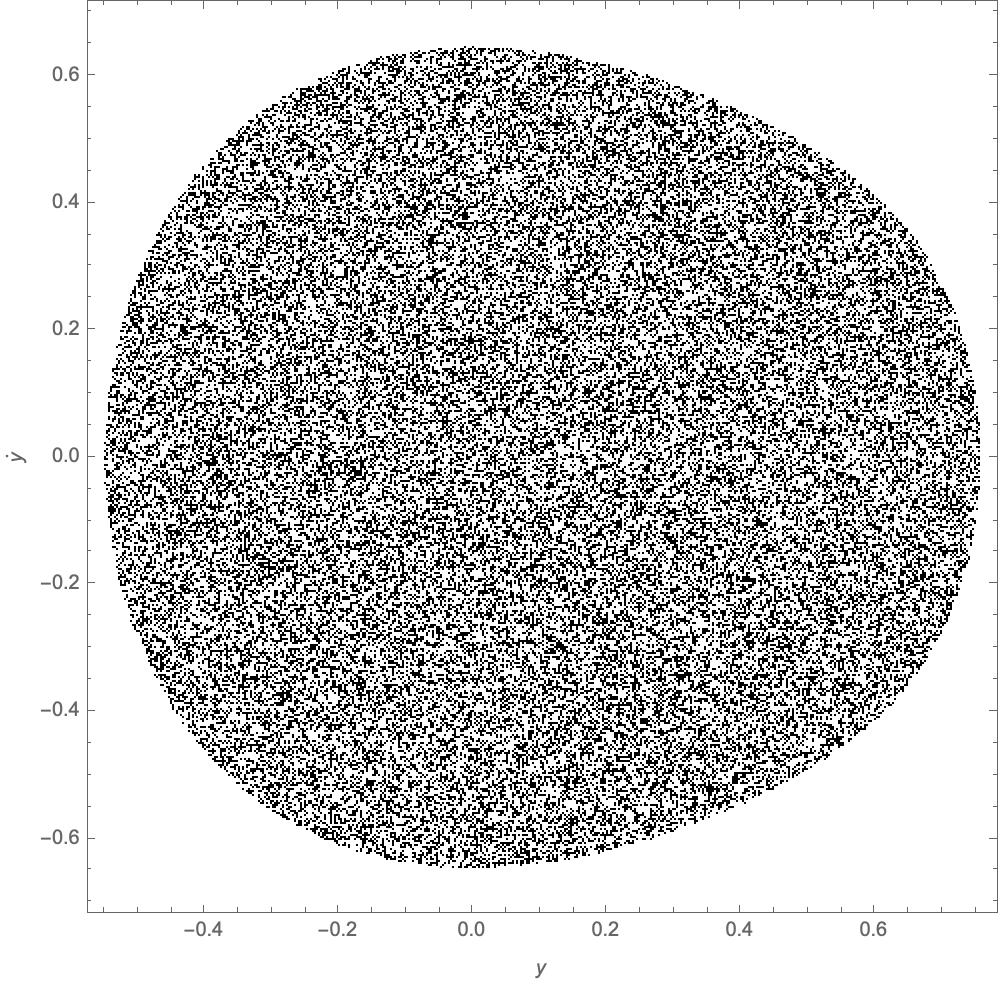}
    }
    \subfloat[]{
        \includegraphics[width=0.3\textwidth]{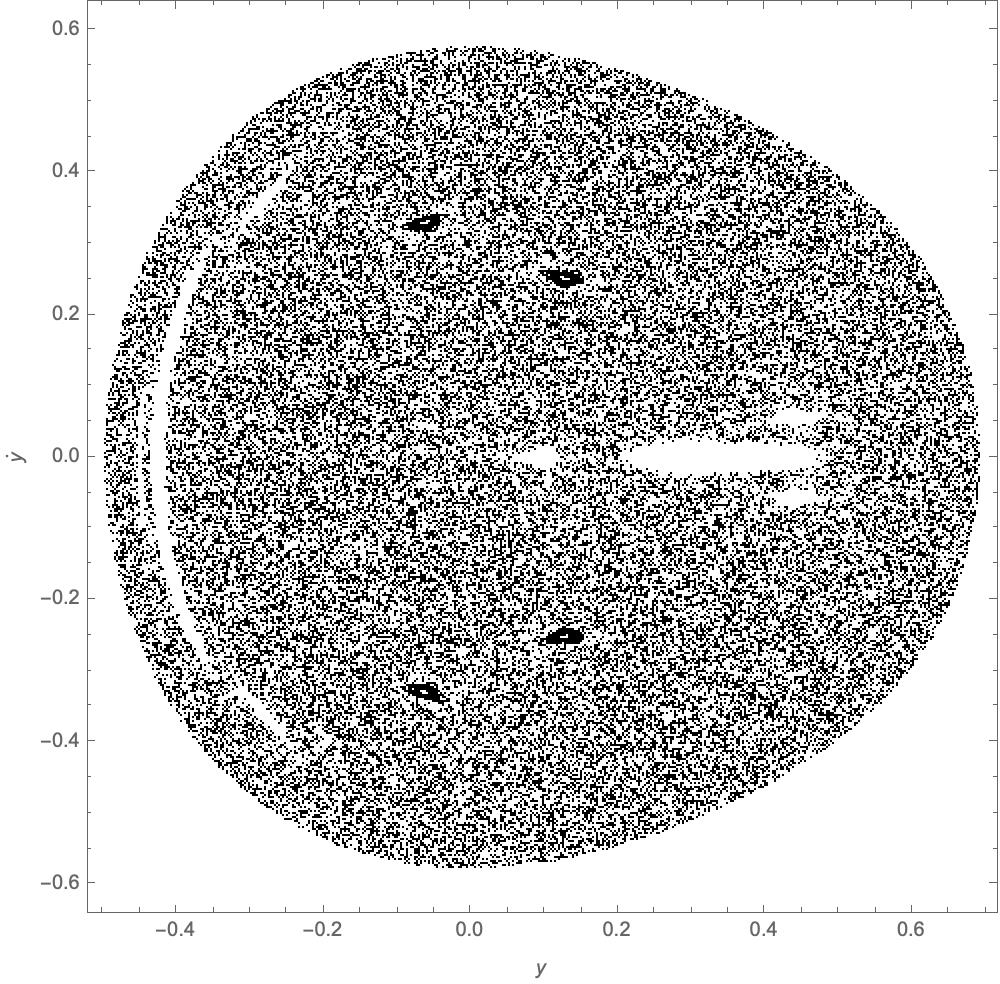}
    }
    \subfloat[]{
        \includegraphics[width=0.3\textwidth]{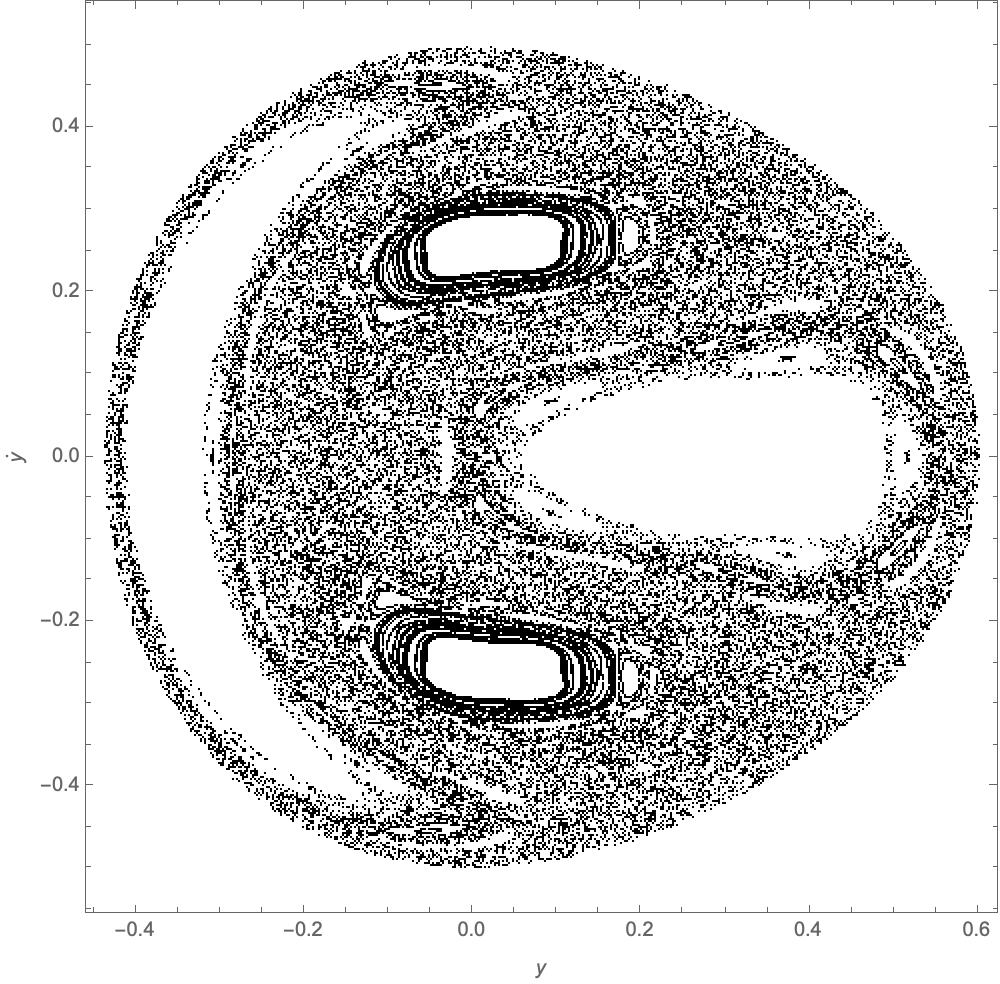}
    }
    \hfill
    \subfloat[]{
        \includegraphics[width=0.3\textwidth]{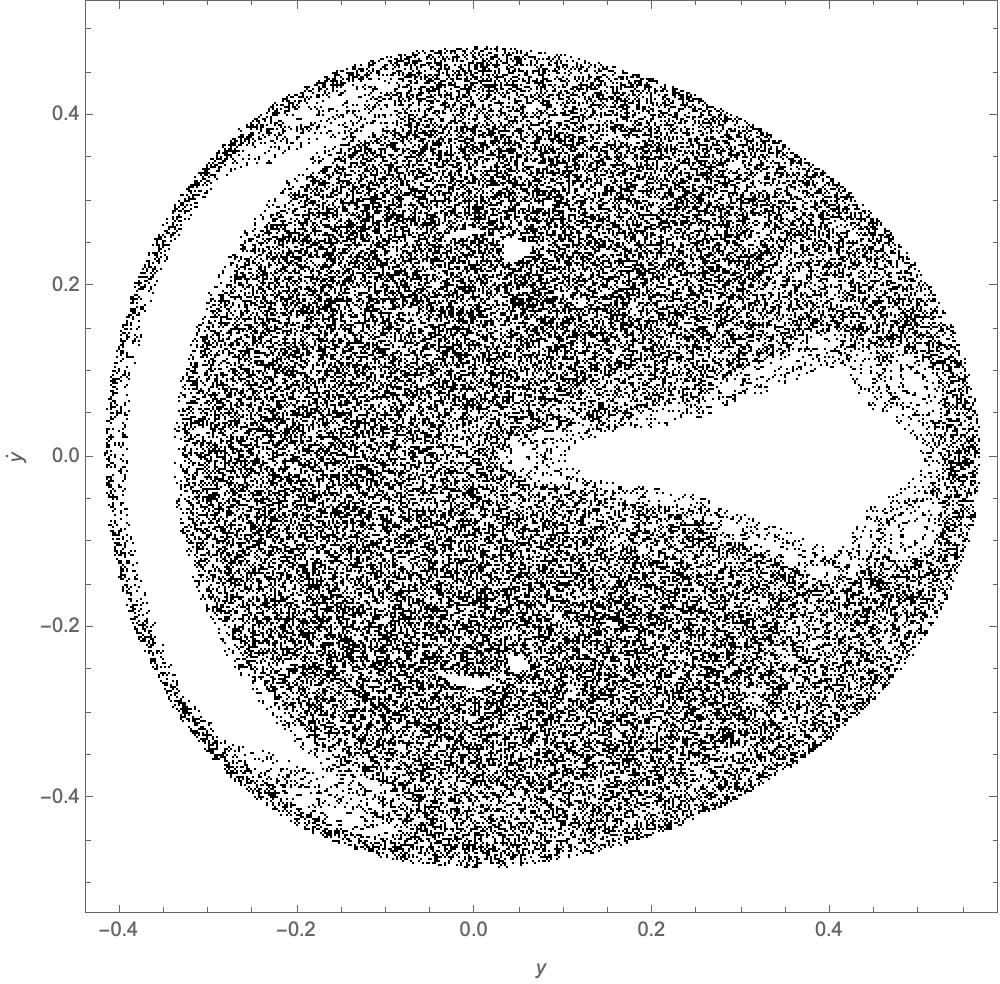}
    }
    \subfloat[]{
        \includegraphics[width=0.3\textwidth]{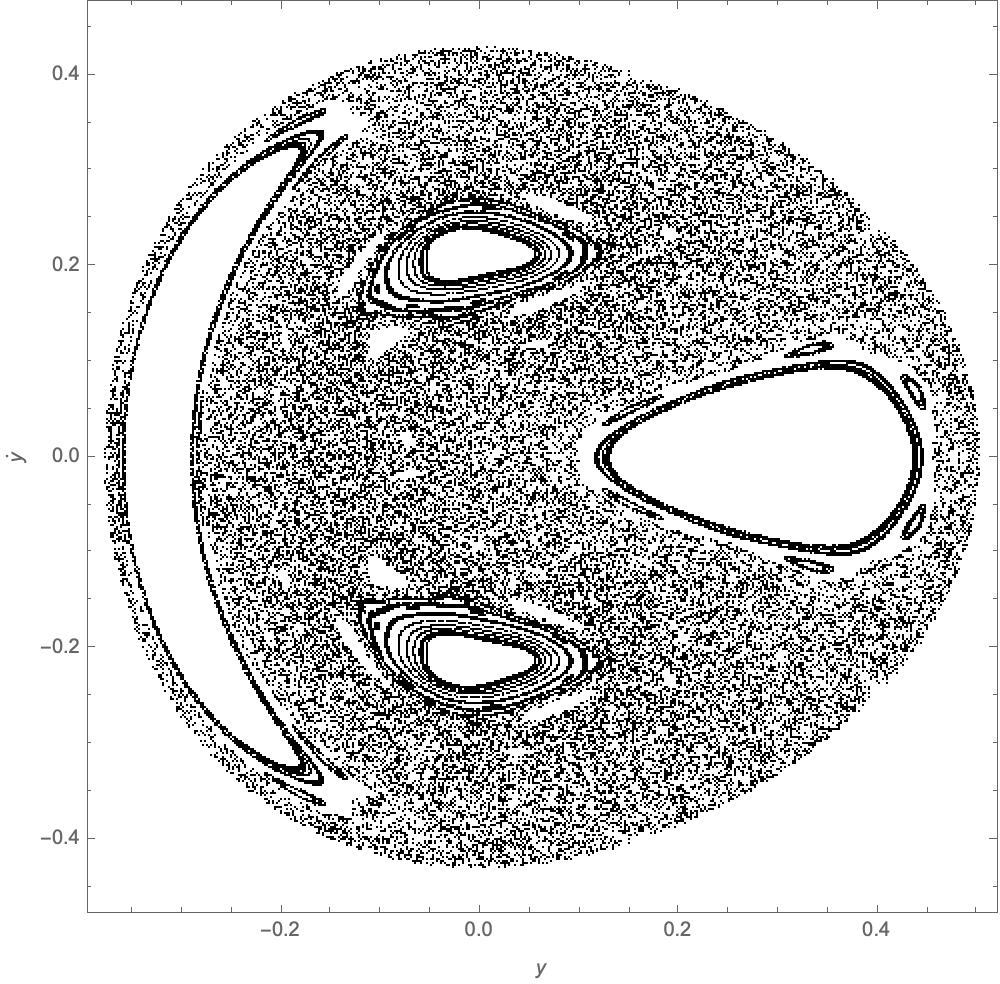}
    }
    \subfloat[]{
        \includegraphics[width=0.3\textwidth]{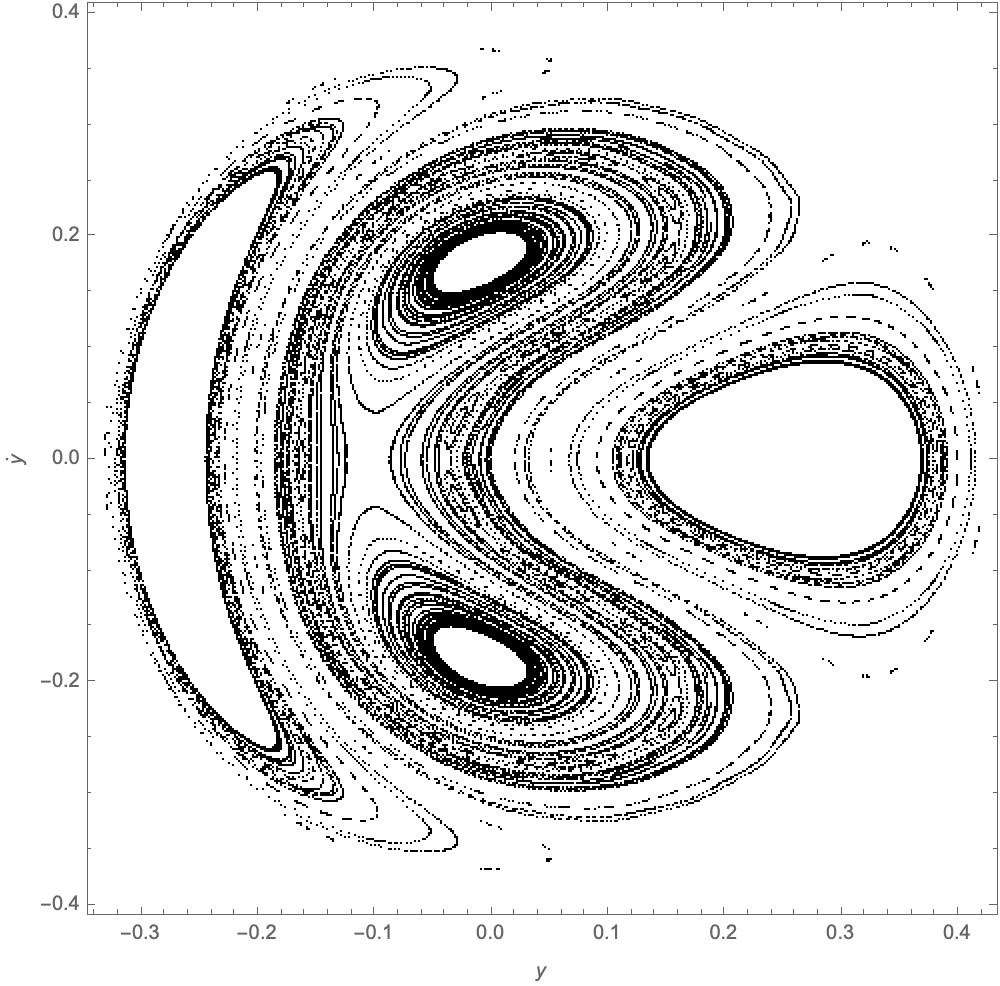}
    }
    \hfill
    \subfloat[]{
        \includegraphics[width=0.3\textwidth]{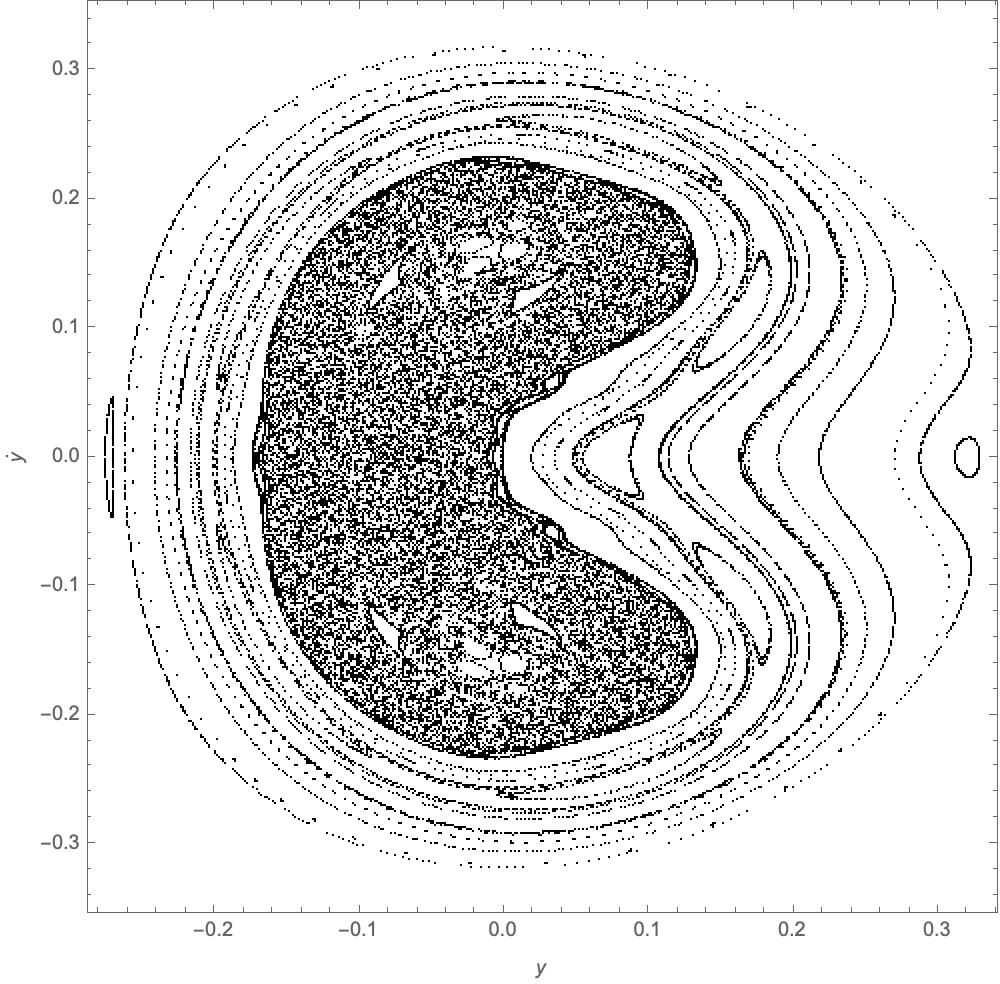}
    }
    \subfloat[]{
        \includegraphics[width=0.3\textwidth]{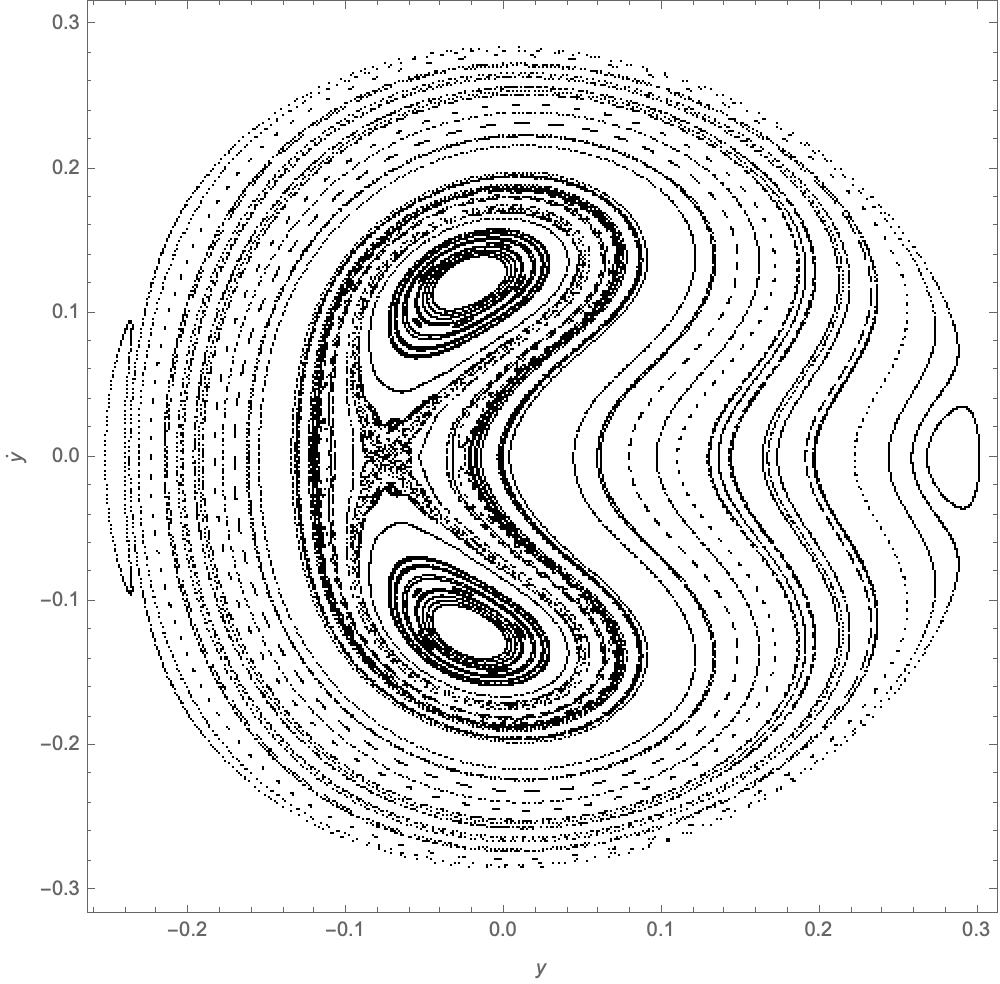}
    }
    \subfloat[]{
        \includegraphics[width=0.3\textwidth]{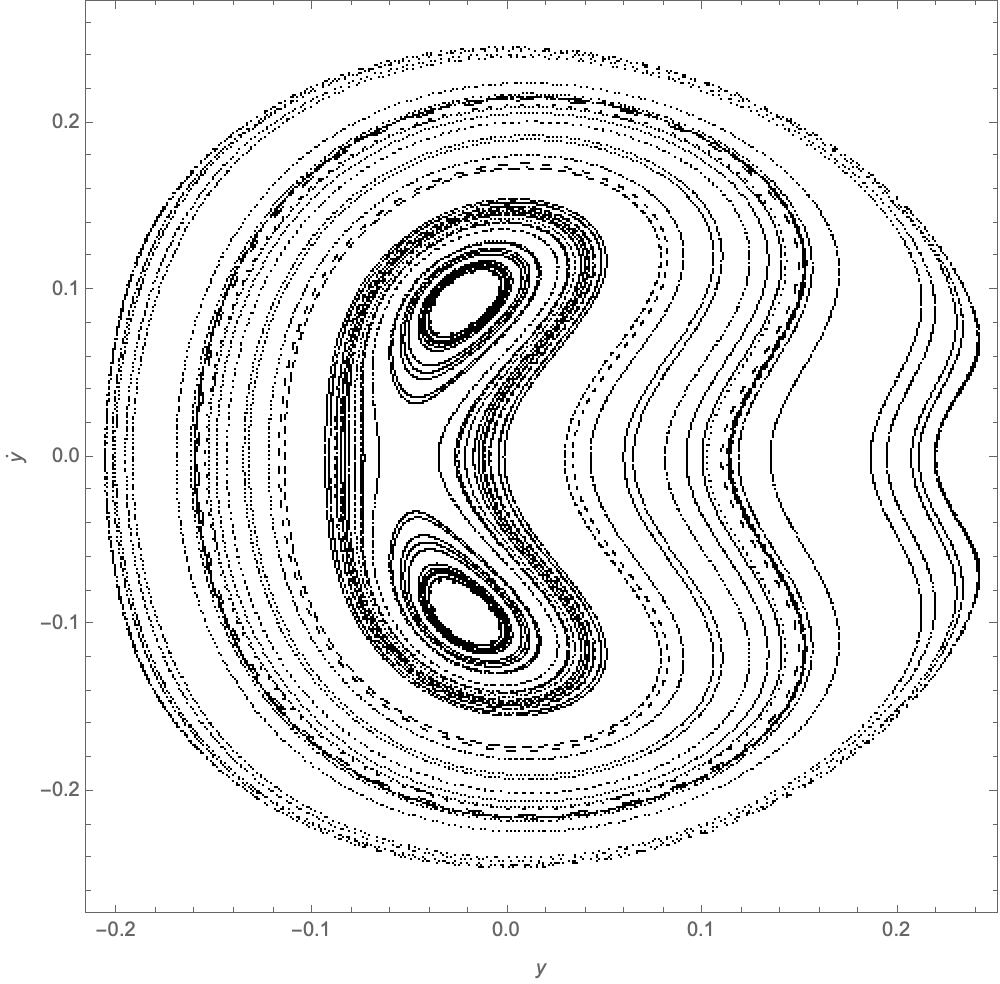}
    }
    \hfill
    \subfloat[]{
        \includegraphics[width=0.3\textwidth]{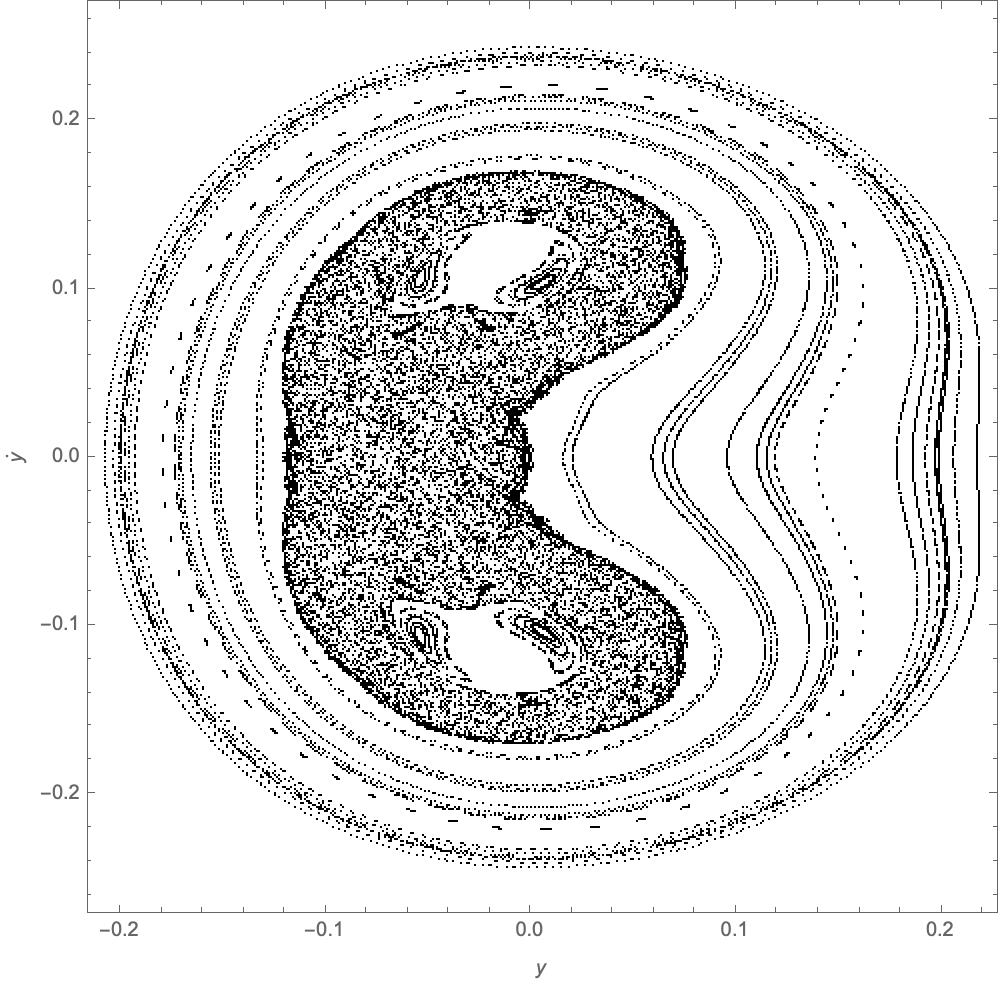}
    }
    \subfloat[]{
        \includegraphics[width=0.3\textwidth]{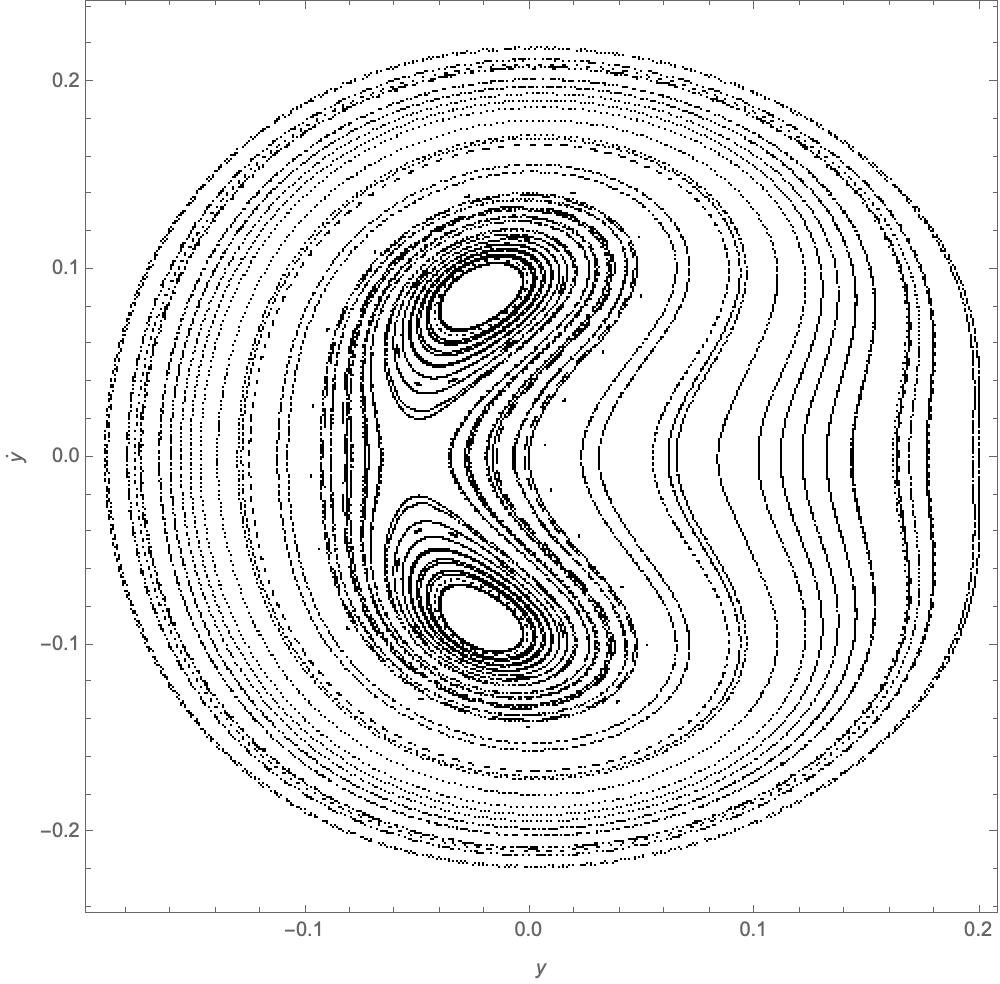}
    }
    \subfloat[   ]{
        \includegraphics[width=0.3\textwidth]{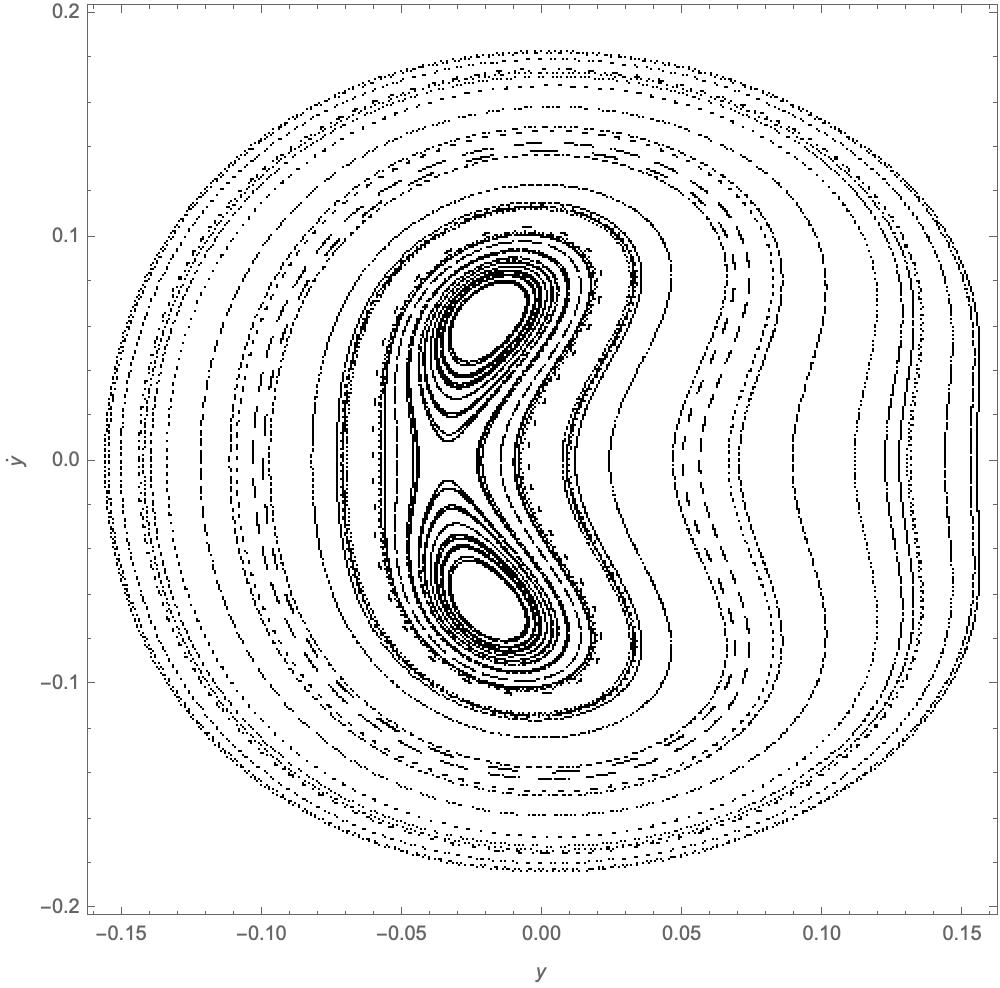}
    }
    \caption{Poincaré sections when $\alpha = 0.1$ and (a) $\delta = 0$ and $n = 1$, (b) $\delta = 0$ and $n = 21$, (c) $\delta = 0$ and $n = 41$, (d) $\delta = 0.1$ and $n = 1$, (e) $\delta = 0.1$ and $n = 21$, (f) $\delta = 0.1$ and $n = 41$, (g) $\delta = 0.5$ and $n = 1$, (h) $\delta = 0.5$ and $n = 21$, (i) $\delta = 0.5$ and $n = 41$, (j) $\delta = 1.0$ and $n = 1$, (k) $\delta = 1.0$ and $n = 21$, (l) $\delta = 1.0$ and $n = 41$ where the energy for each panel is $E = E_{\text{min}}(1-n/100)$.}
    \label{fig:alpha=0.1}
\end{figure}

\begin{figure}[htbp]
    \centering
    \subfloat[]{
        \includegraphics[width=0.3\textwidth]{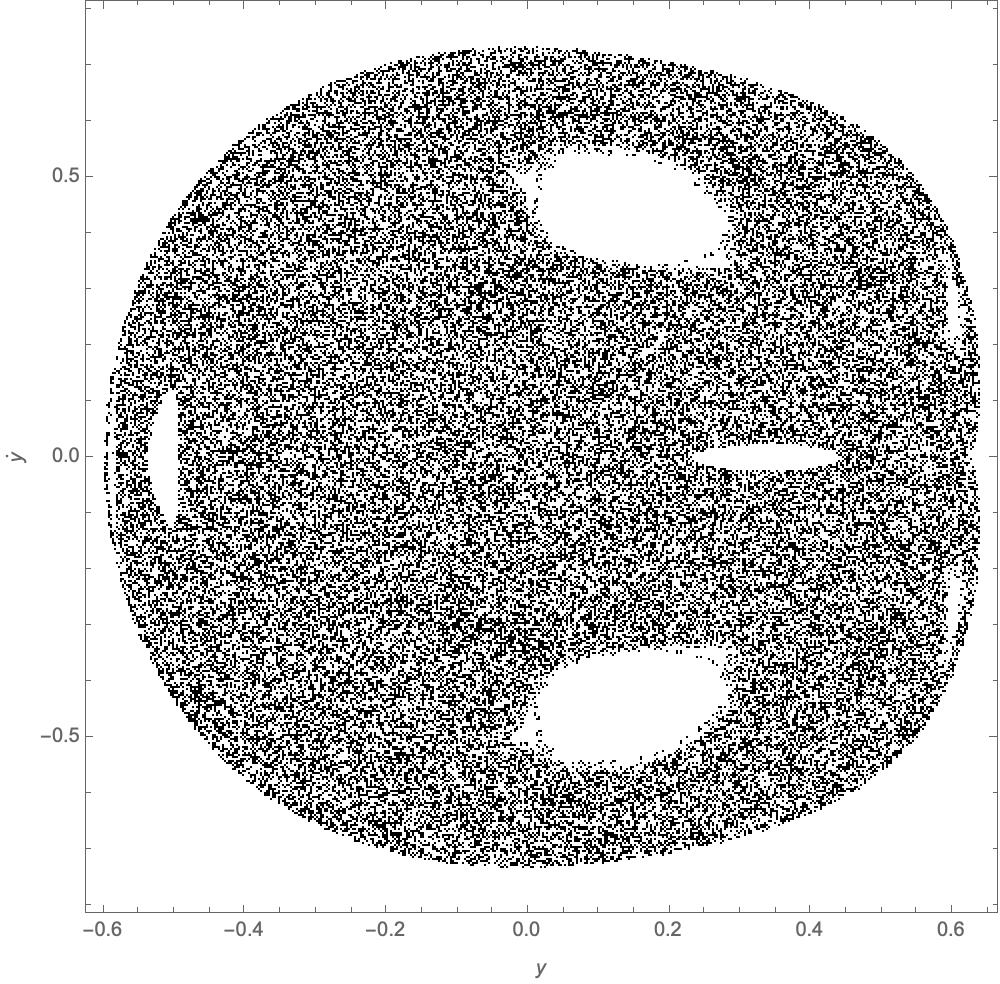}
    }
    \subfloat[]{
        \includegraphics[width=0.3\textwidth]{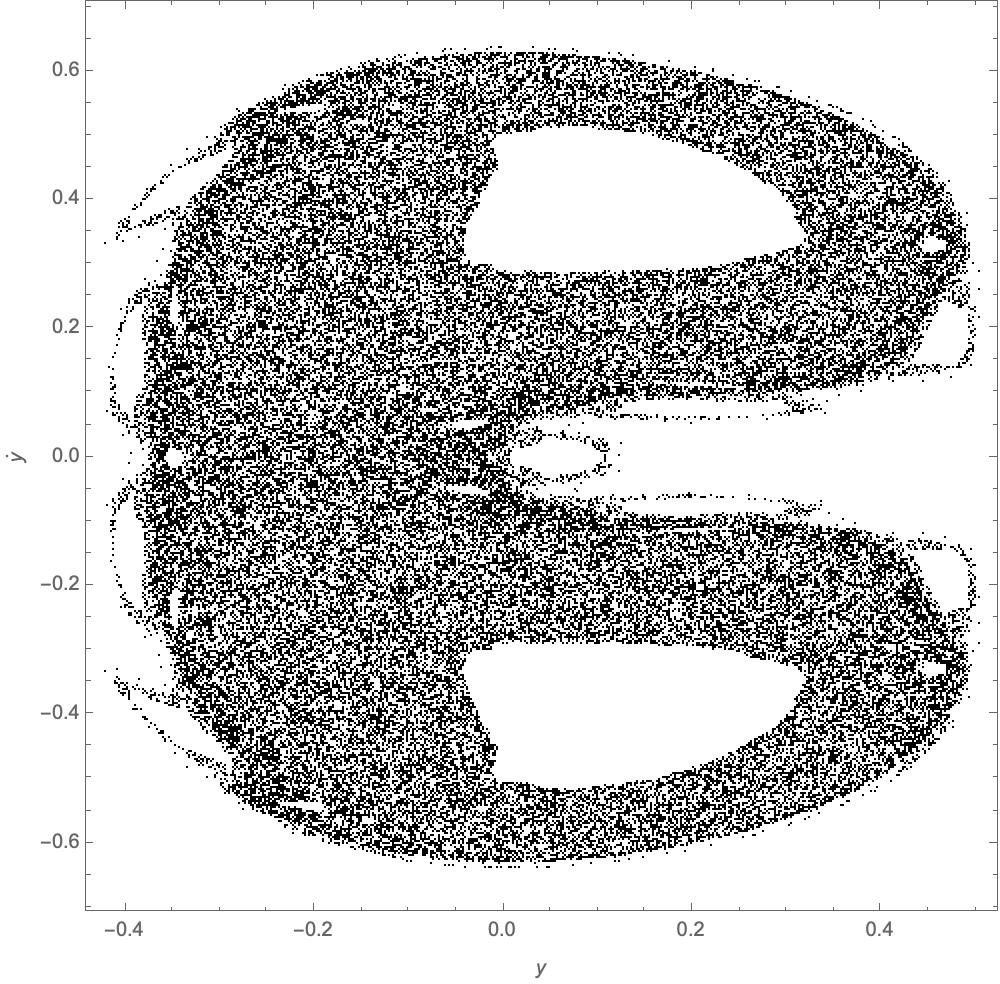}
    }
    \subfloat[]{
        \includegraphics[width=0.3\textwidth]{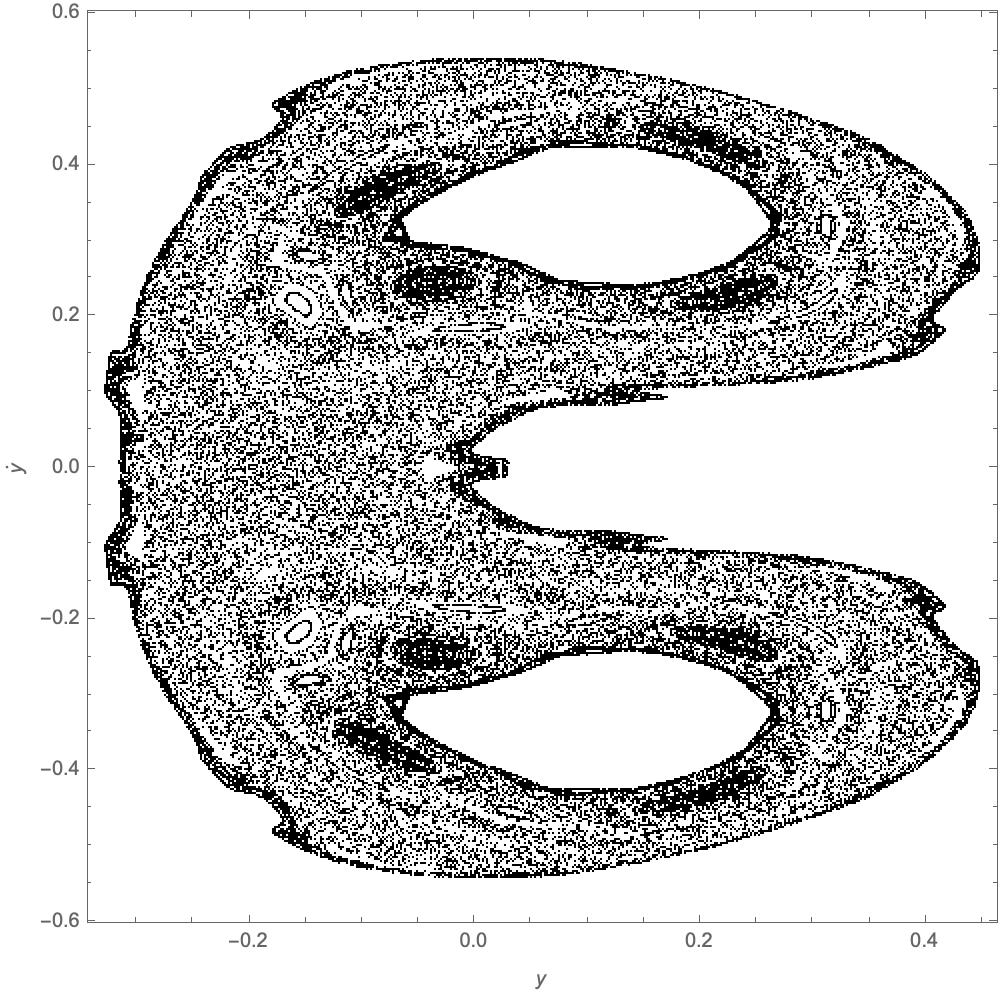}
    }
    \hfill
    \subfloat[]{
        \includegraphics[width=0.3\textwidth]{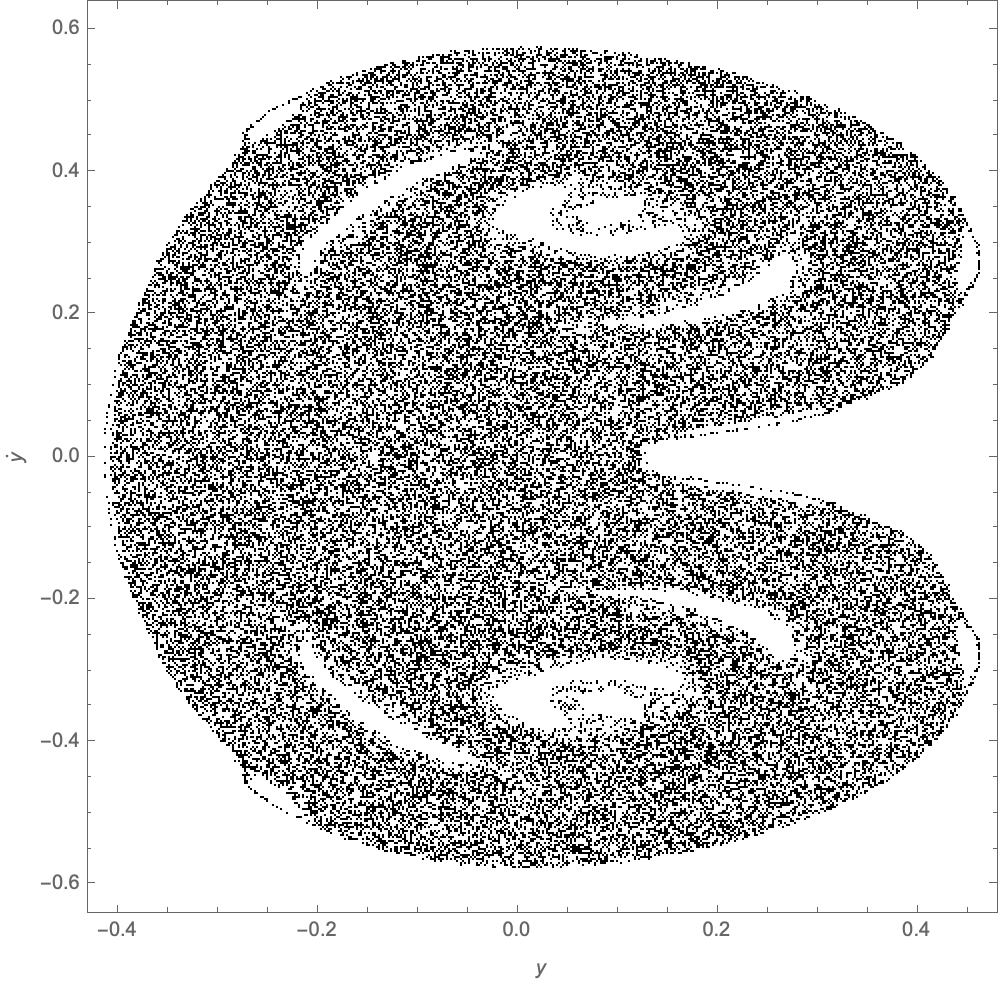}
    }
    \subfloat[]{
        \includegraphics[width=0.3\textwidth]{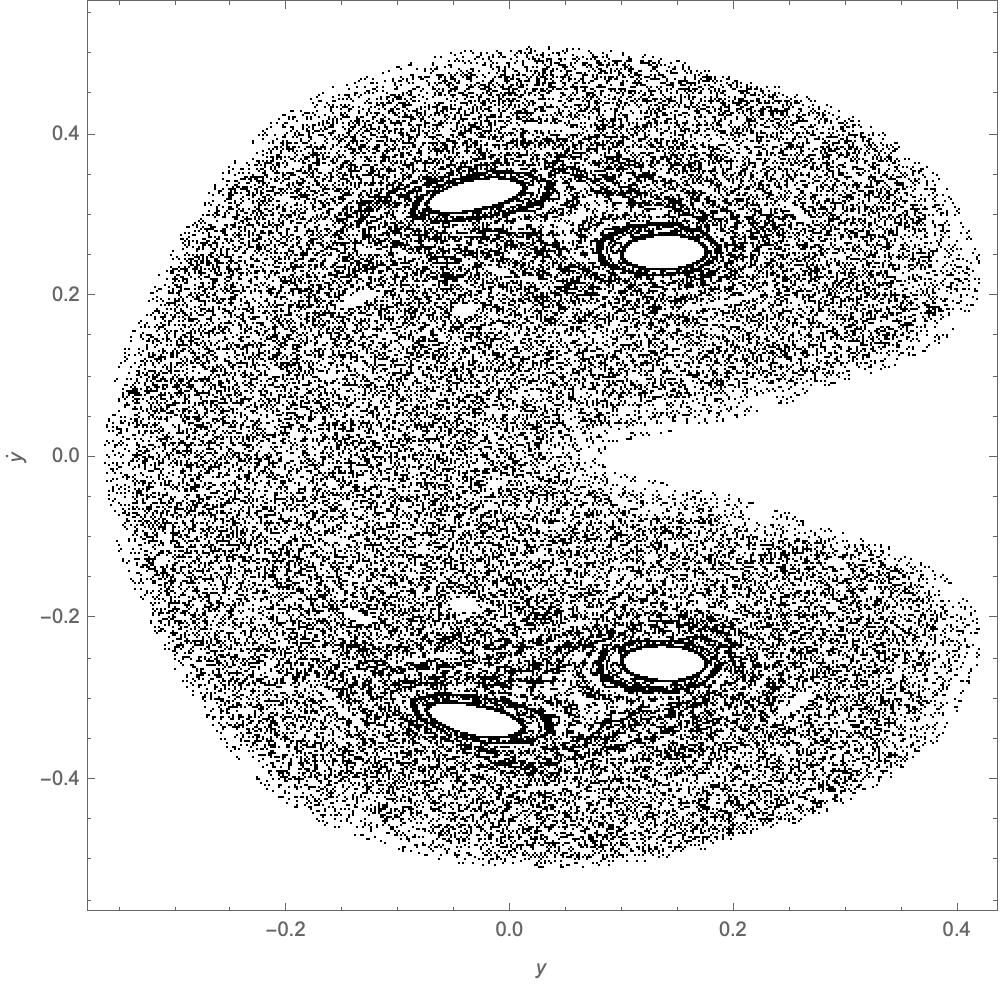}
    }
    \subfloat[]{
        \includegraphics[width=0.3\textwidth]{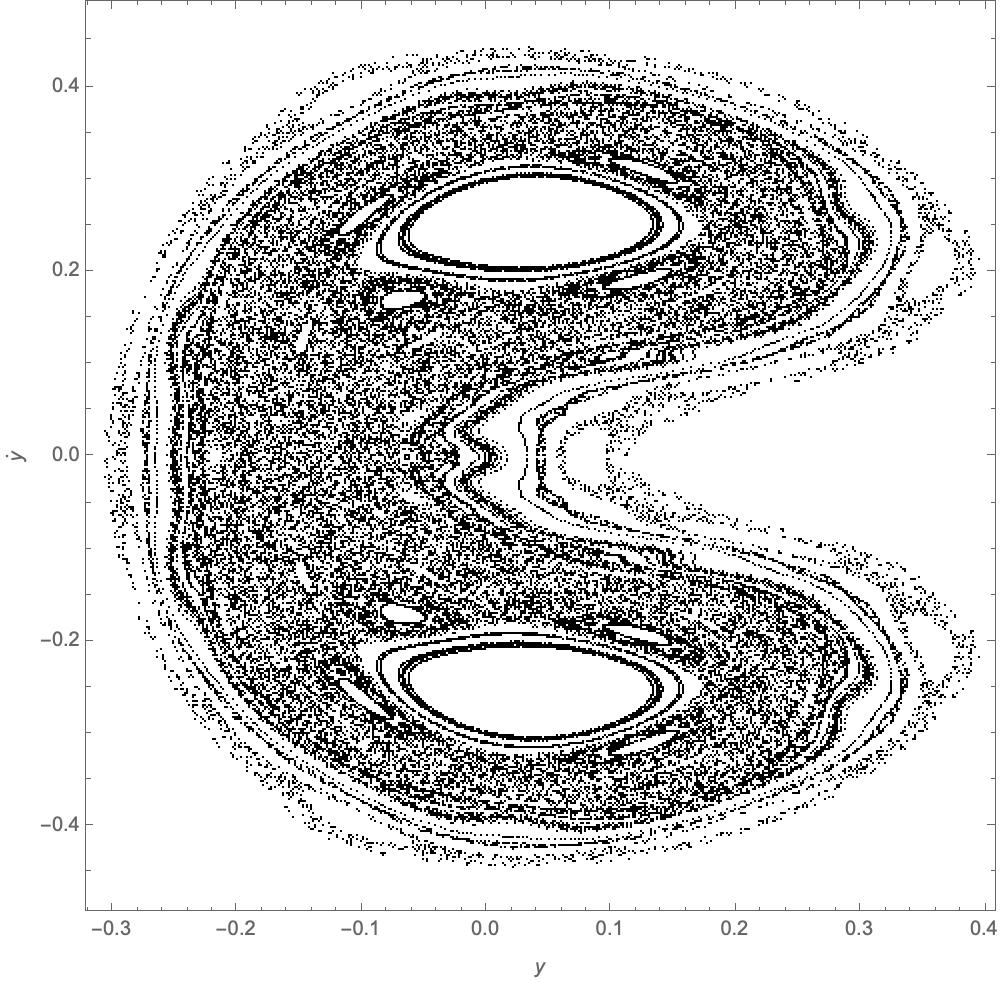}
    }
    \hfill
    \subfloat[]{
        \includegraphics[width=0.3\textwidth]{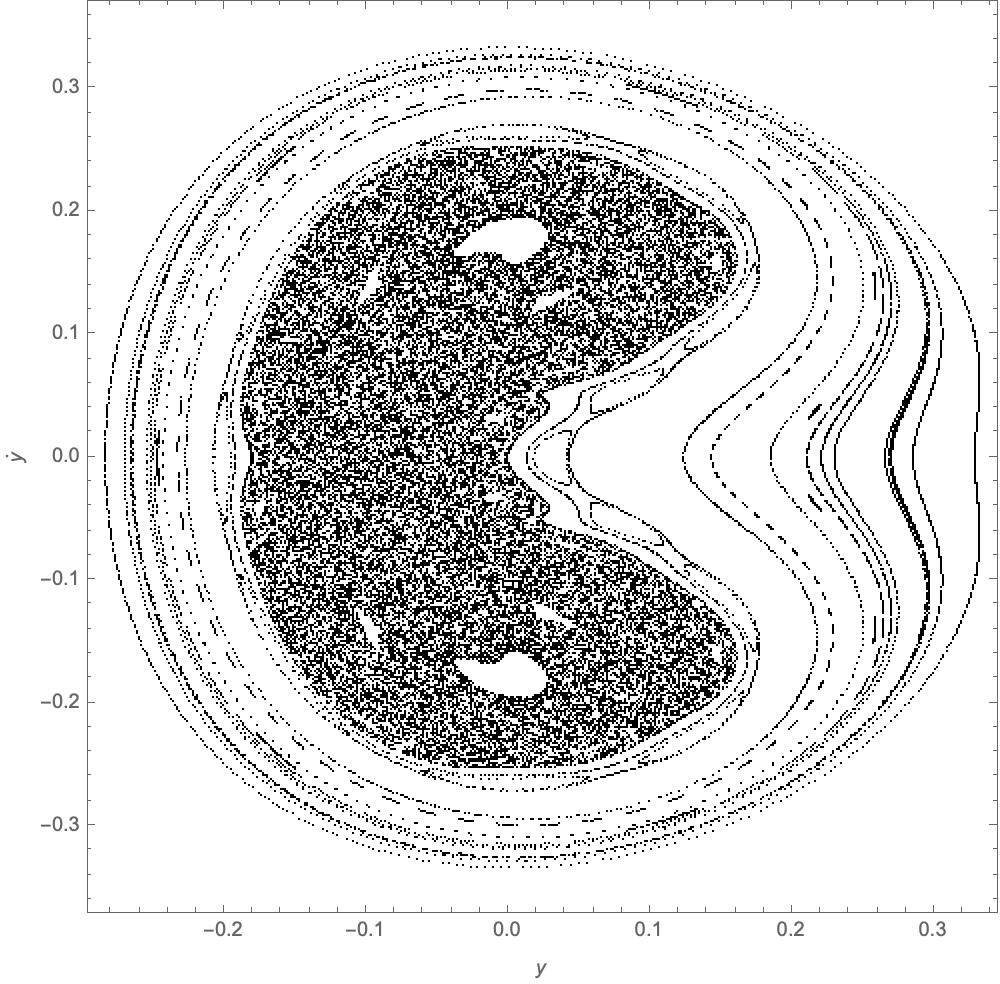}
    }
    \subfloat[]{
        \includegraphics[width=0.3\textwidth]{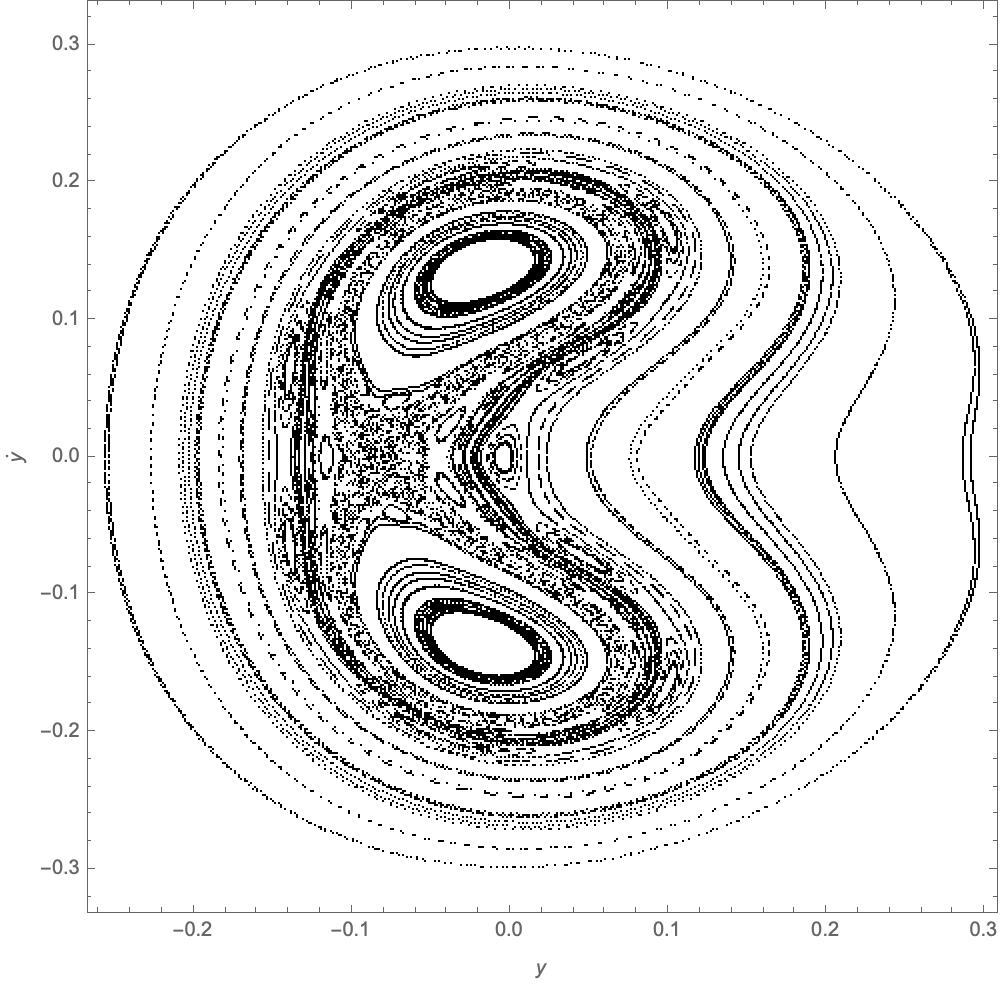}
    }
    \subfloat[]{
        \includegraphics[width=0.3\textwidth]{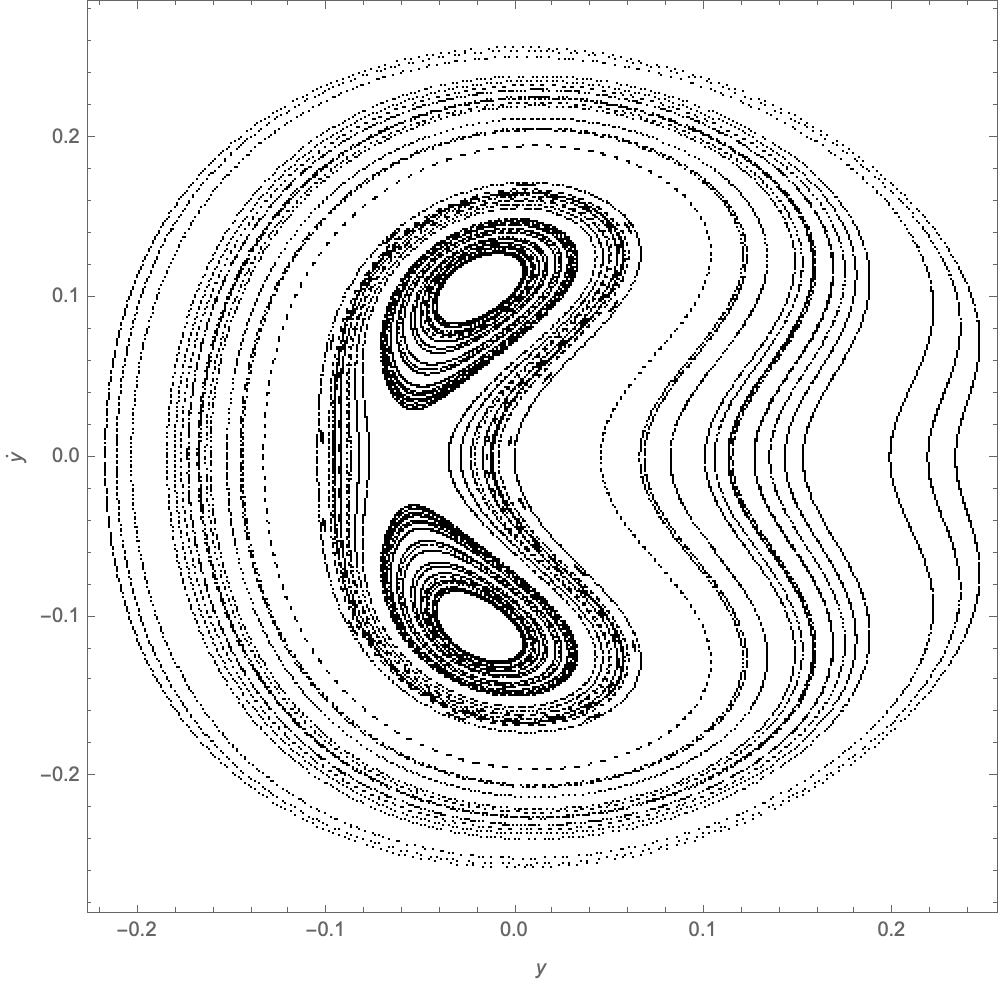}
    }
    \hfill
    \subfloat[]{
        \includegraphics[width=0.3\textwidth]{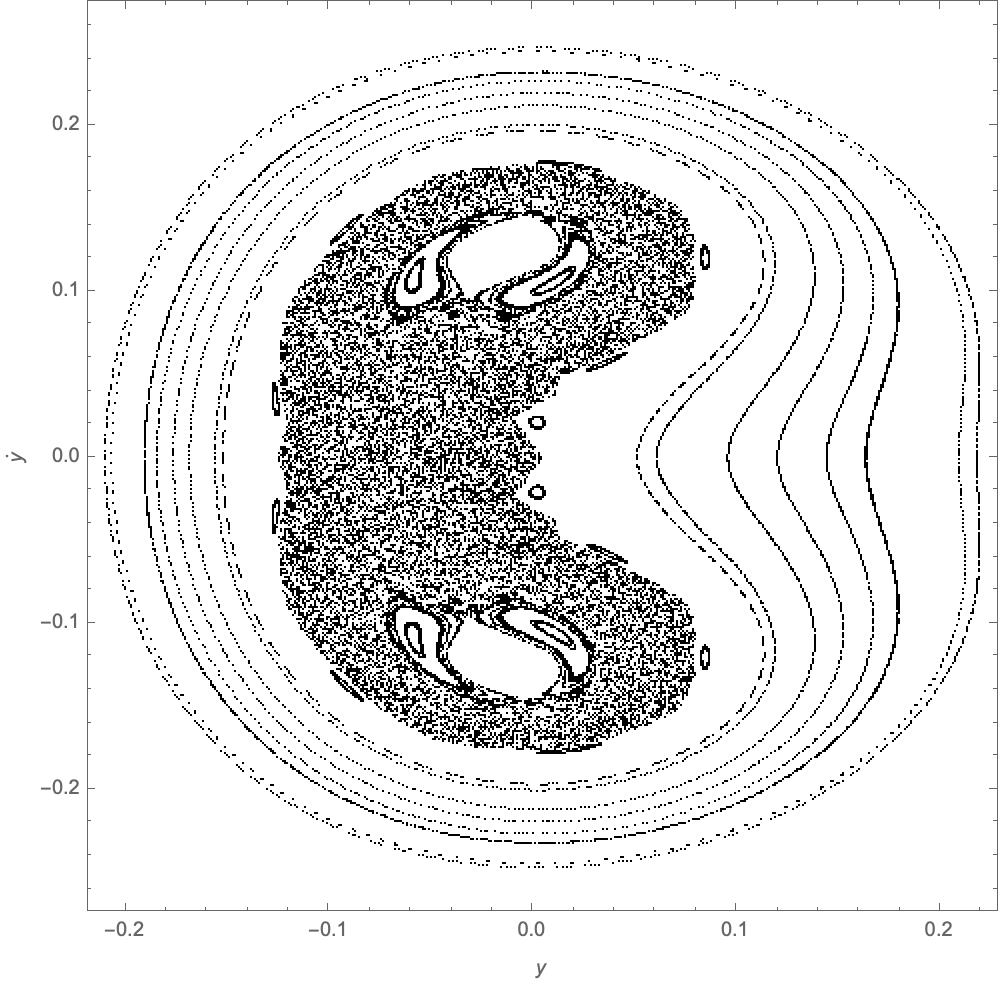}
    }
    \subfloat[]{
        \includegraphics[width=0.3\textwidth]{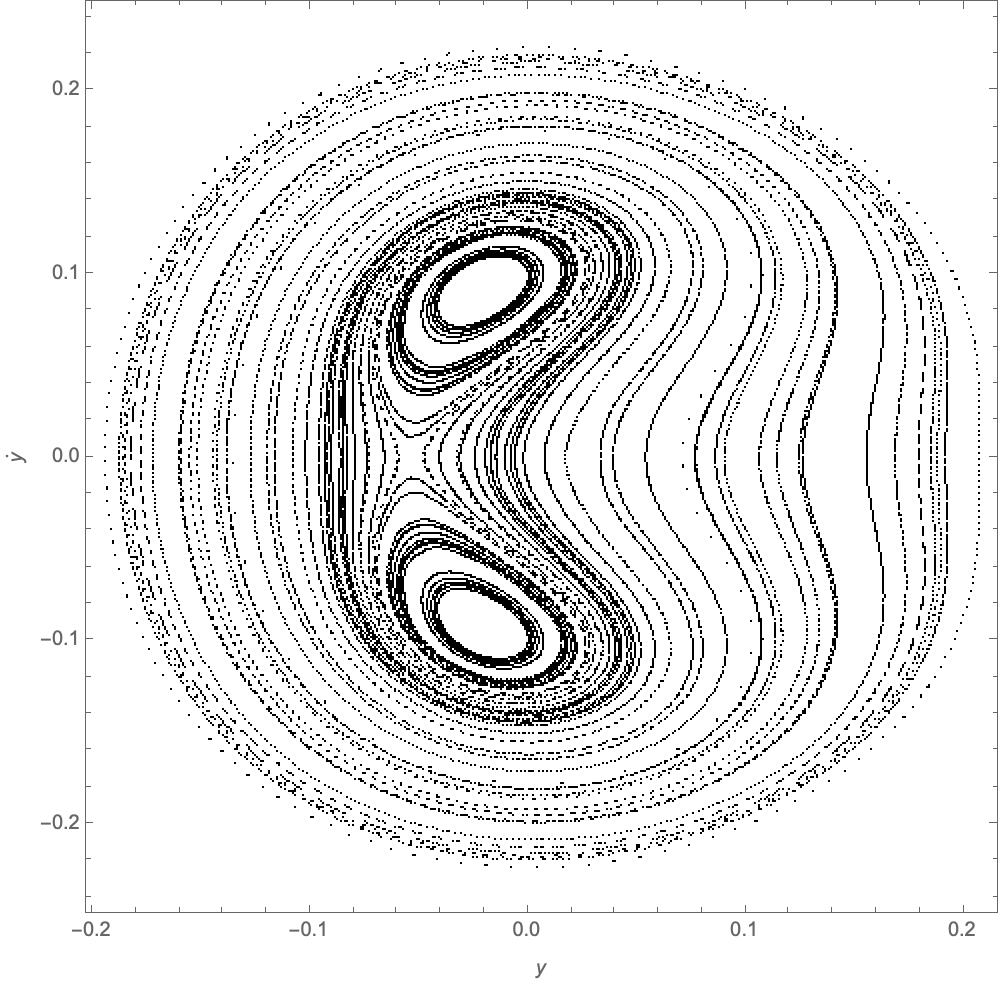}
    }
    \subfloat[   ]{
        \includegraphics[width=0.3\textwidth]{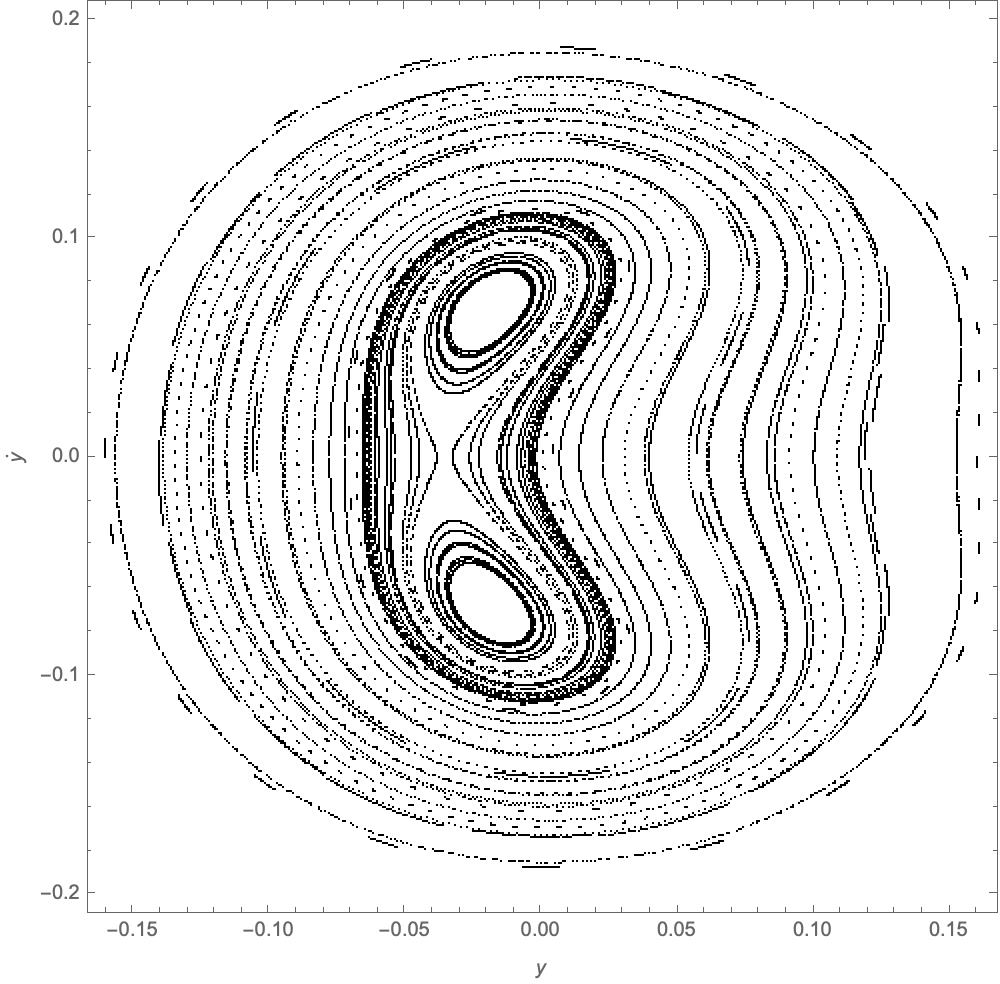}
    }
    \caption{Poincaré sections when $\alpha = 0.5$ and (a) $\delta = 0$ and $n = 1$, (b) $\delta = 0$ and $n = 21$, (c) $\delta = 0$ and $n = 41$, (d) $\delta = 0.1$ and $n = 1$, (e) $\delta = 0.1$ and $n = 21$, (f) $\delta = 0.1$ and $n = 41$, (g) $\delta = 0.5$ and $n = 1$, (h) $\delta = 0.5$ and $n = 21$, (i) $\delta = 0.5$ and $n = 41$, (j) $\delta = 1.0$ and $n = 1$, (k) $\delta = 1.0$ and $n = 21$, (l) $\delta = 1.0$ and $n = 41$ where the energy for each panel is $E = E_{\text{min}}(1-n/100)$.}
    \label{fig:alpha=0.5}
\end{figure}

\begin{figure}[htbp]
    \centering
    \subfloat[]{
        \includegraphics[width=0.3\textwidth]{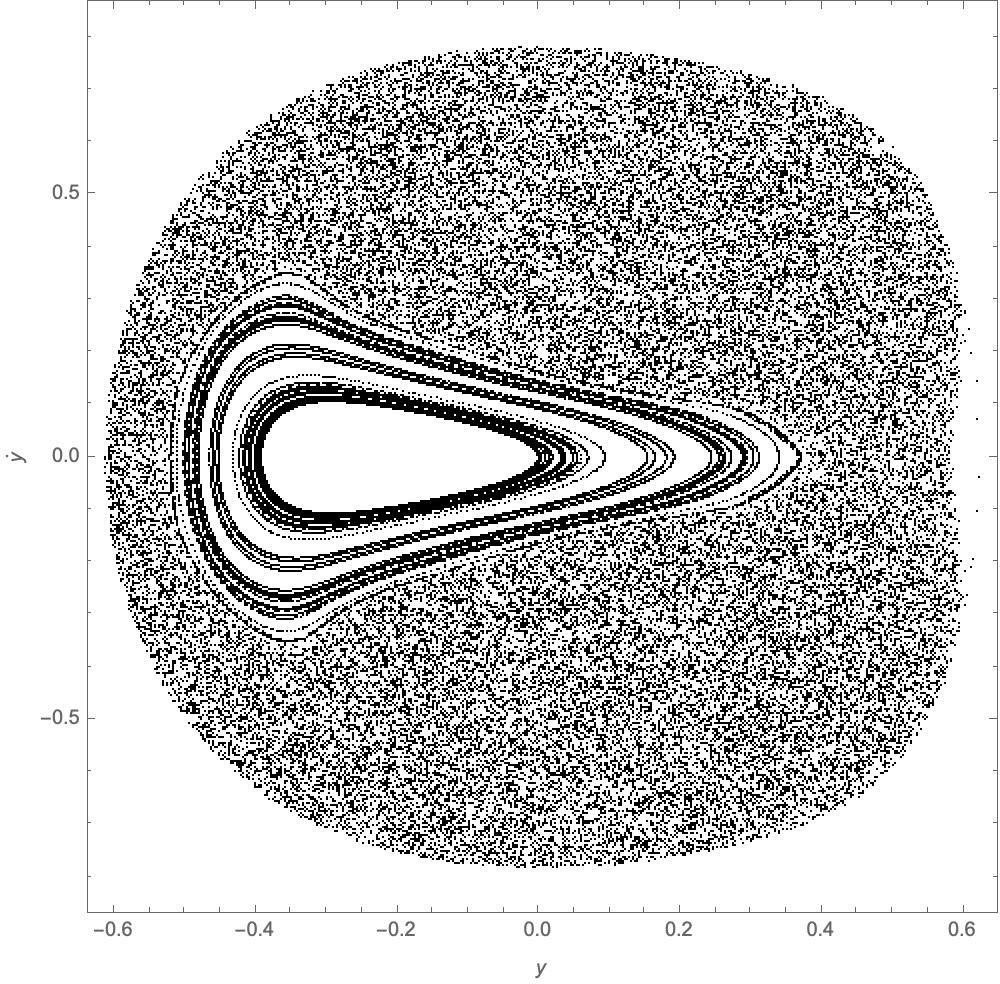}
    }
    \subfloat[]{
        \includegraphics[width=0.3\textwidth]{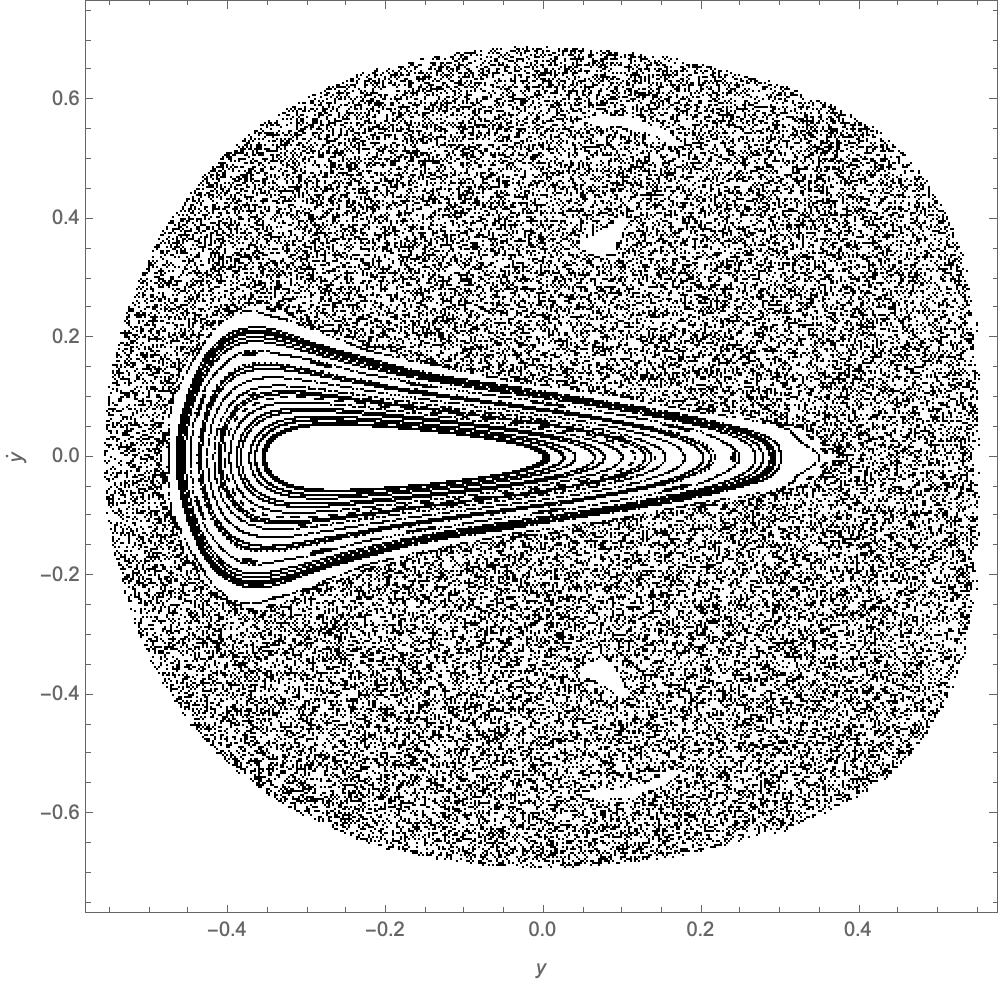}
    }
    \subfloat[]{
        \includegraphics[width=0.3\textwidth]{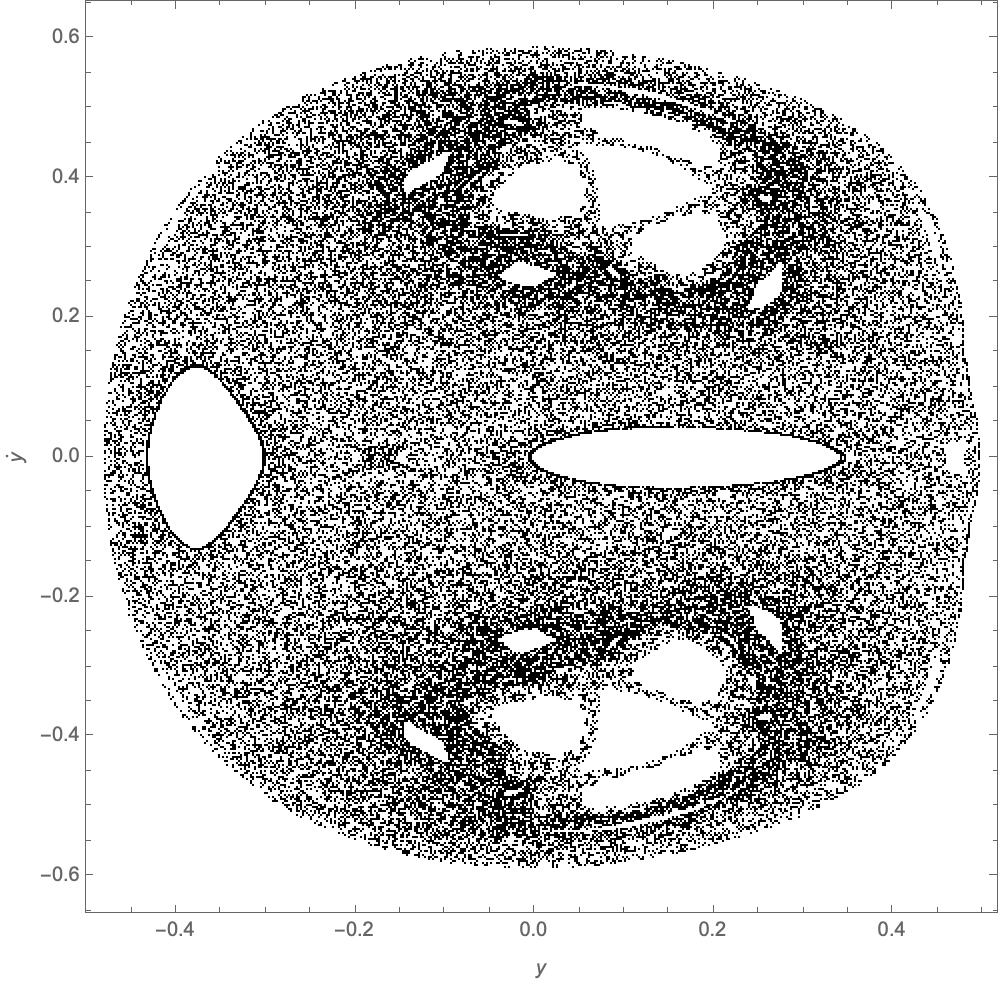}
    }
    \hfill
    \subfloat[]{
        \label{ref_label1}
        \includegraphics[width=0.3\textwidth]{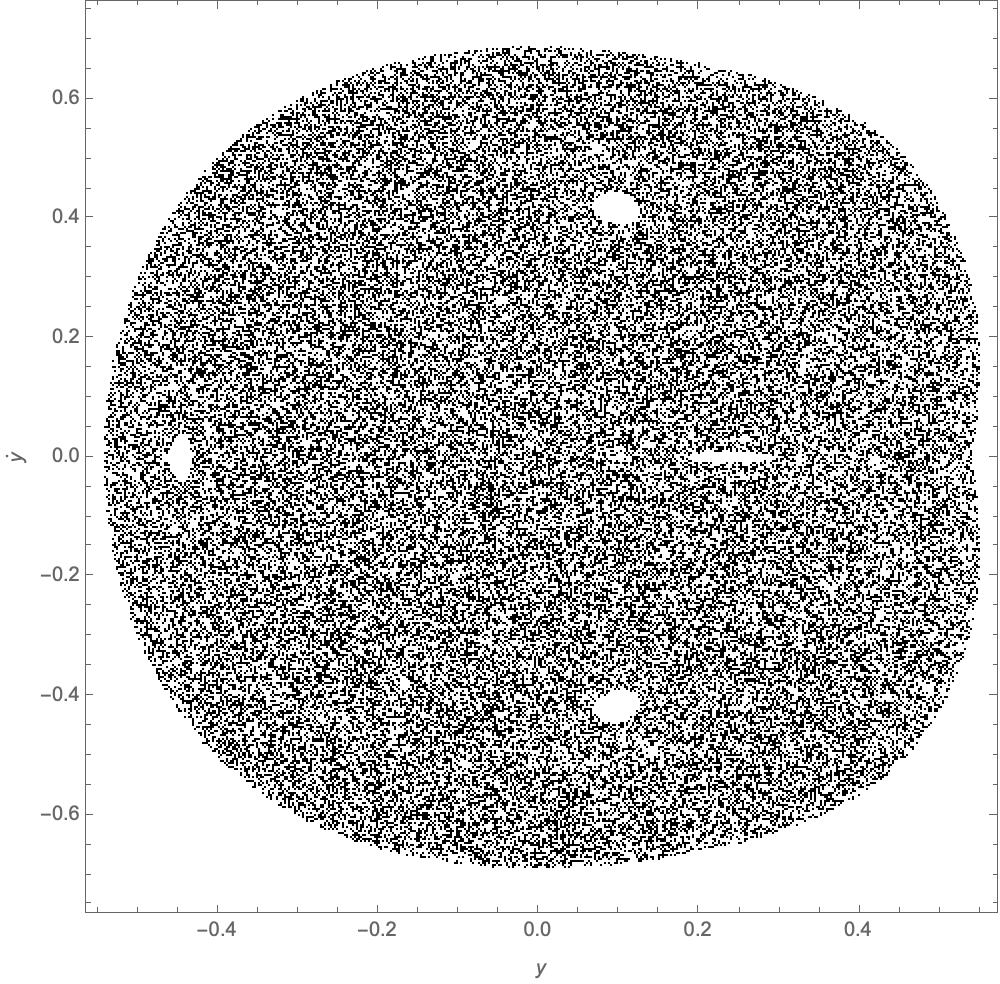}
    }
    \subfloat[]{
        \includegraphics[width=0.3\textwidth]{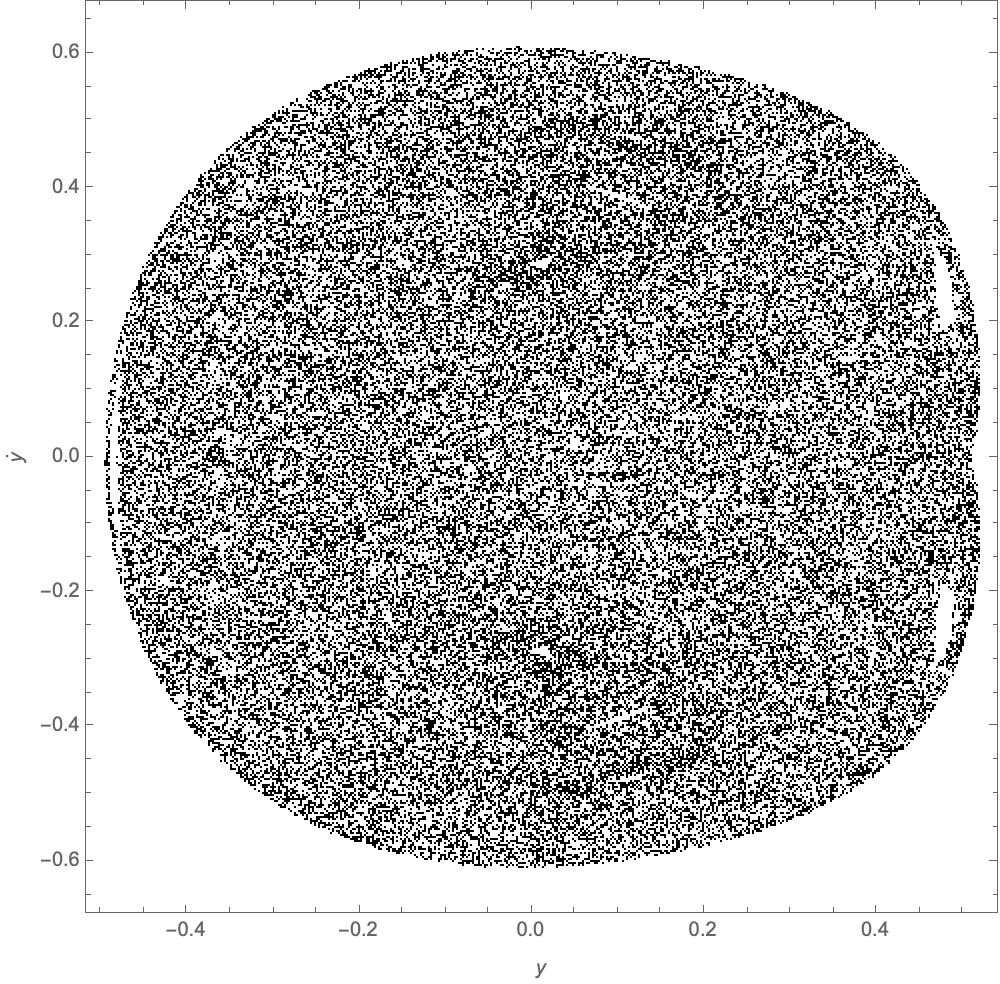}
    }
    \subfloat[]{
        \includegraphics[width=0.3\textwidth]{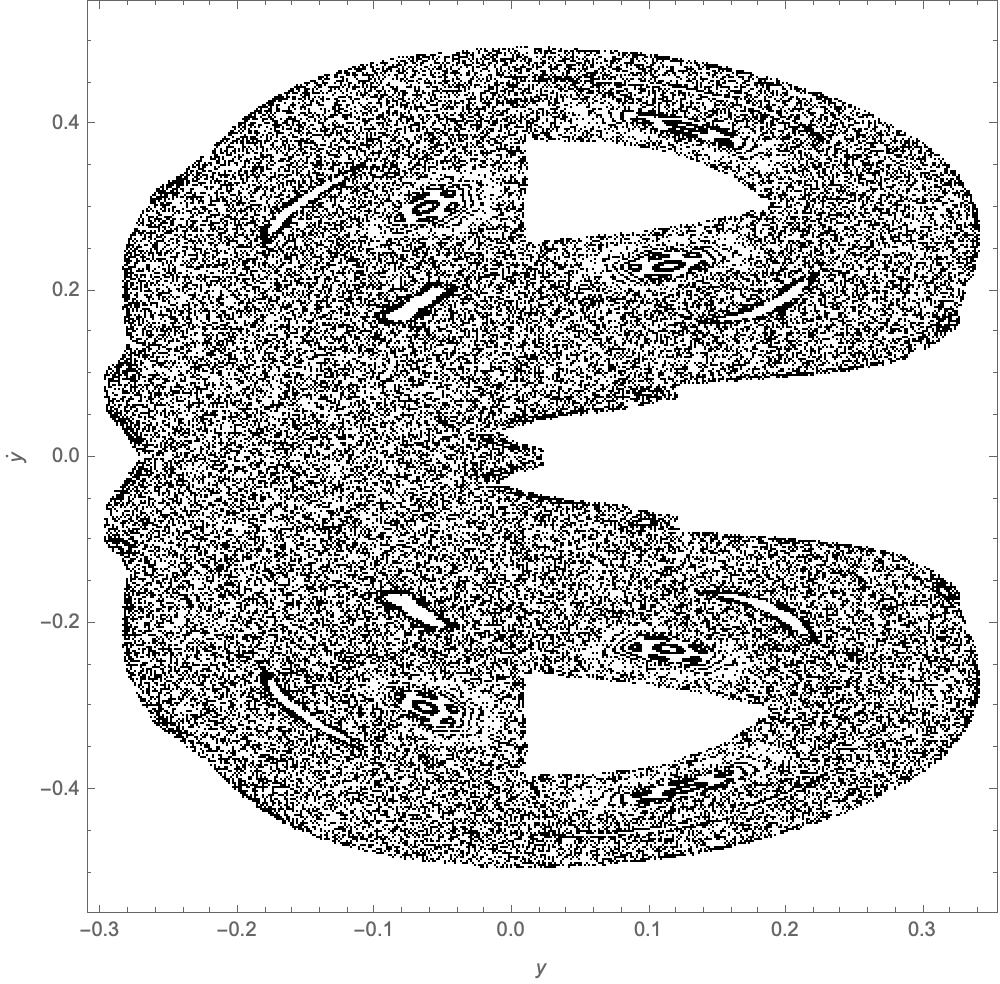}
    }
    \hfill
    \subfloat[]{
        \includegraphics[width=0.3\textwidth]{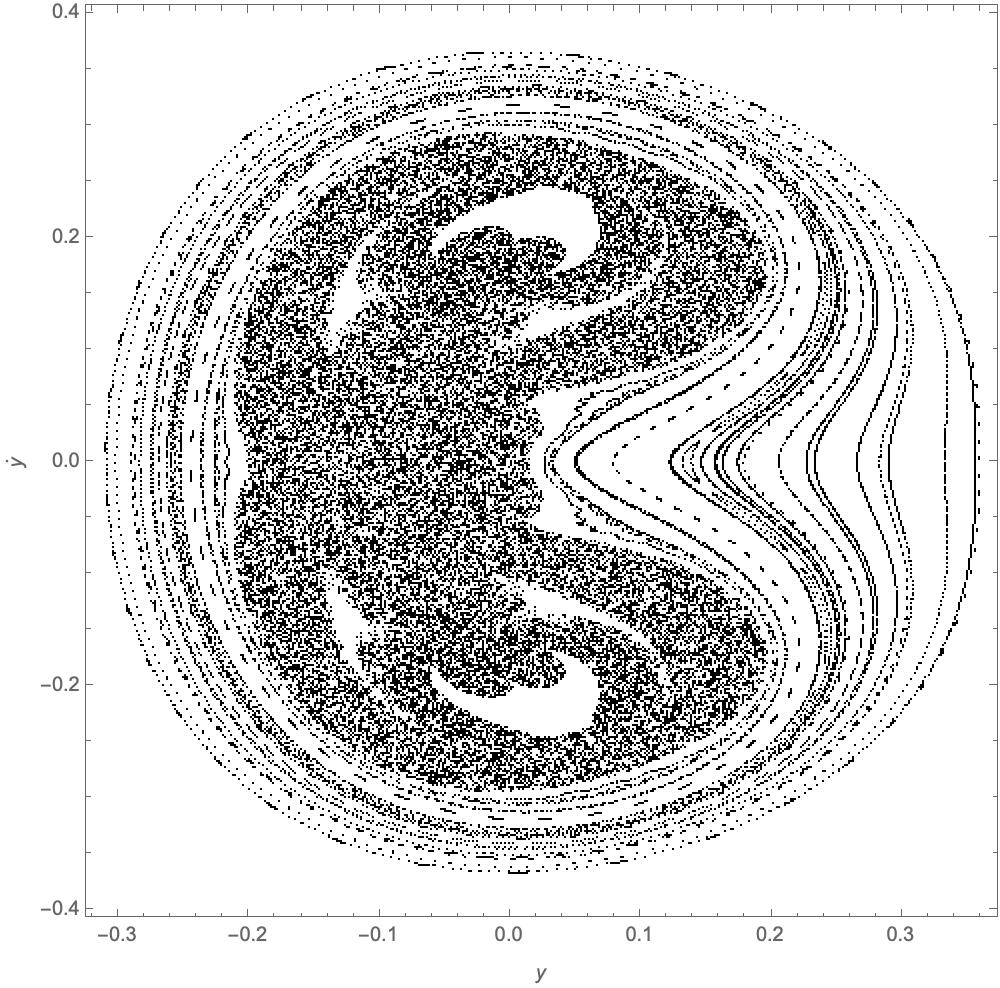}
    }
    \subfloat[]{
        \includegraphics[width=0.3\textwidth]{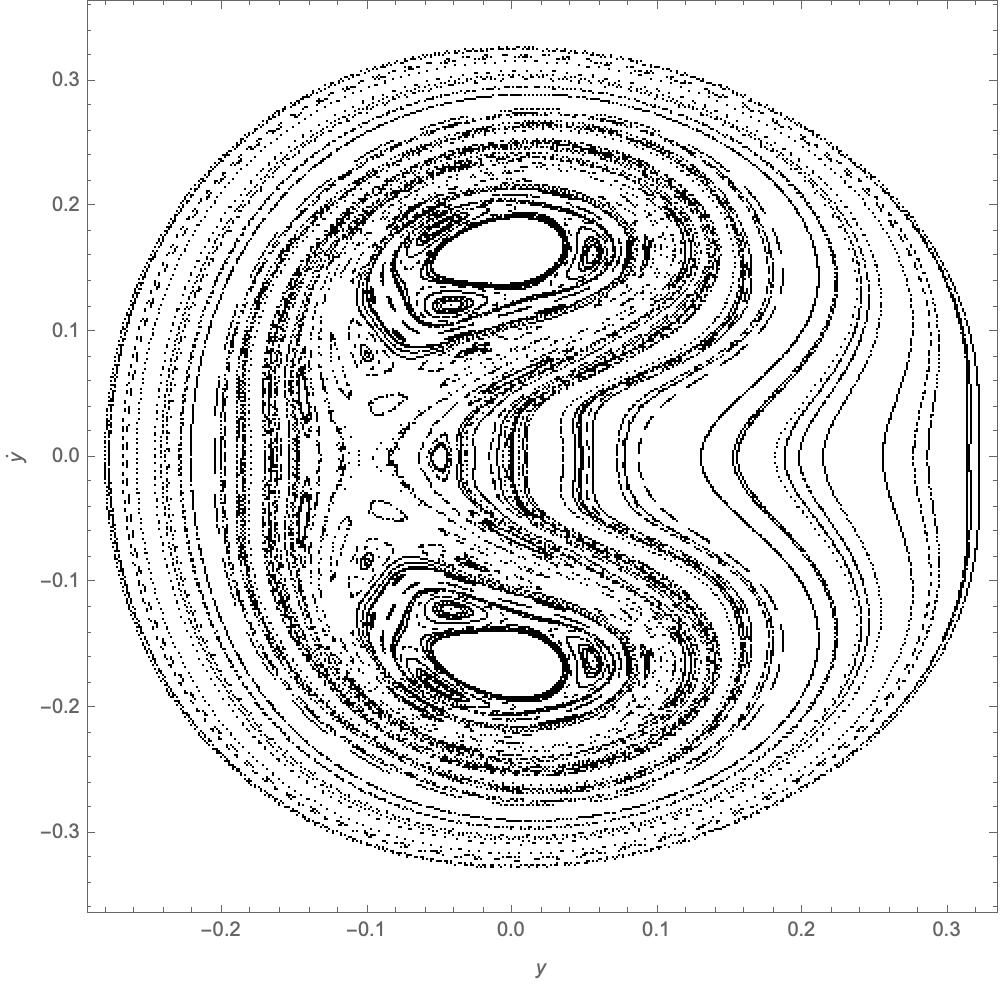}
    }
    \subfloat[]{
        \includegraphics[width=0.3\textwidth]{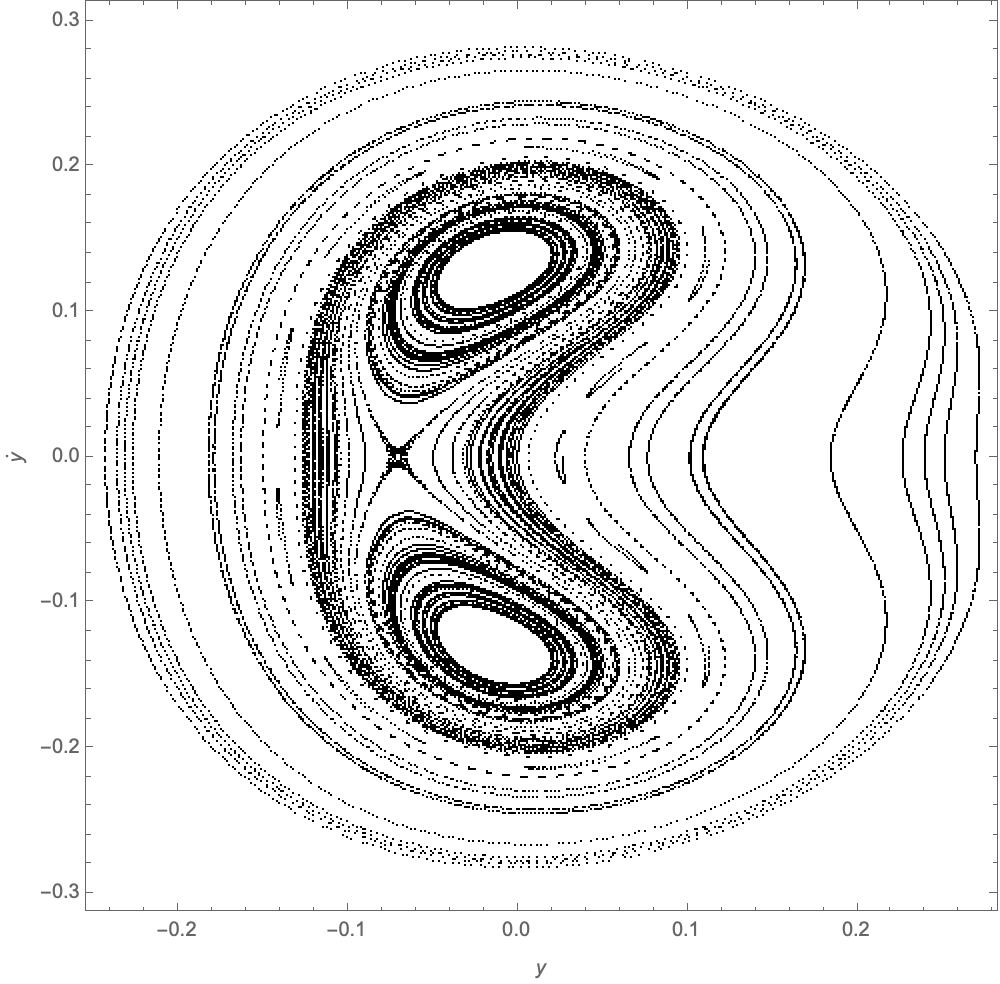}
    }
    \hfill
    \subfloat[]{
        \includegraphics[width=0.3\textwidth]{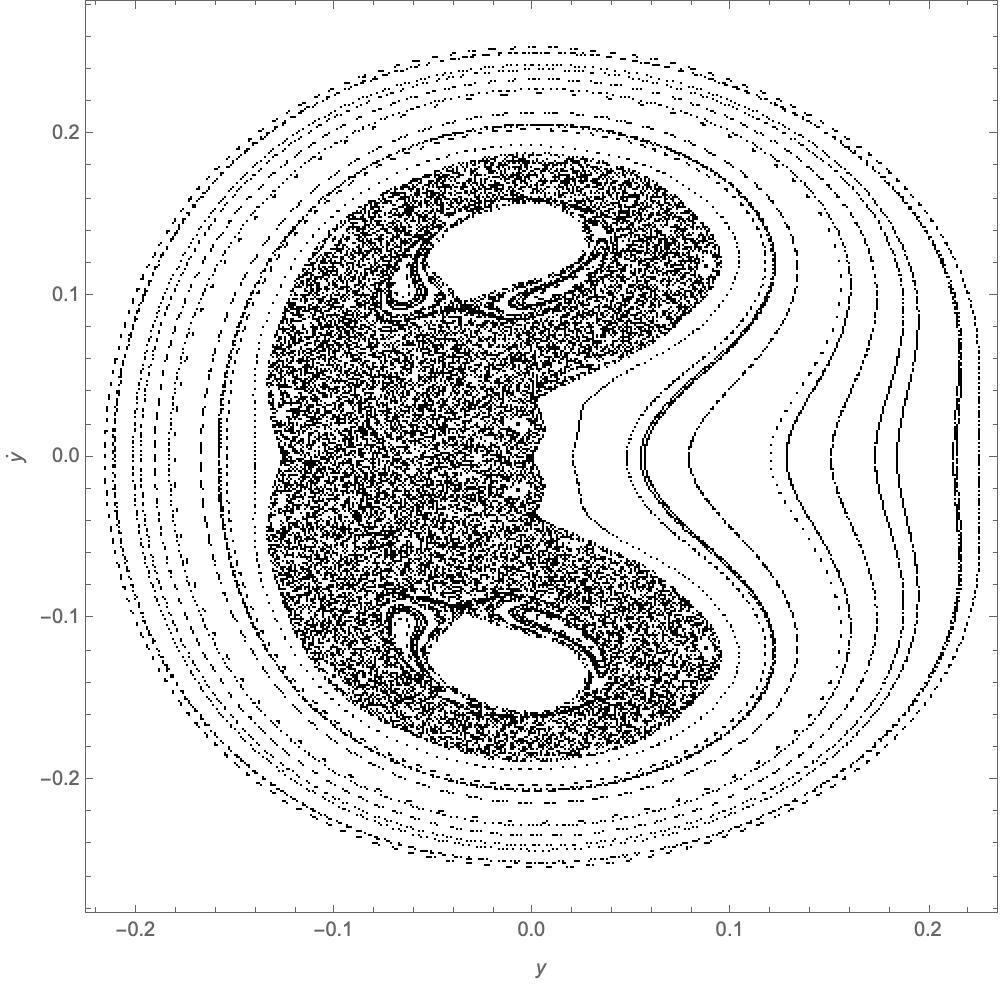}
    }
    \subfloat[]{
        \includegraphics[width=0.3\textwidth]{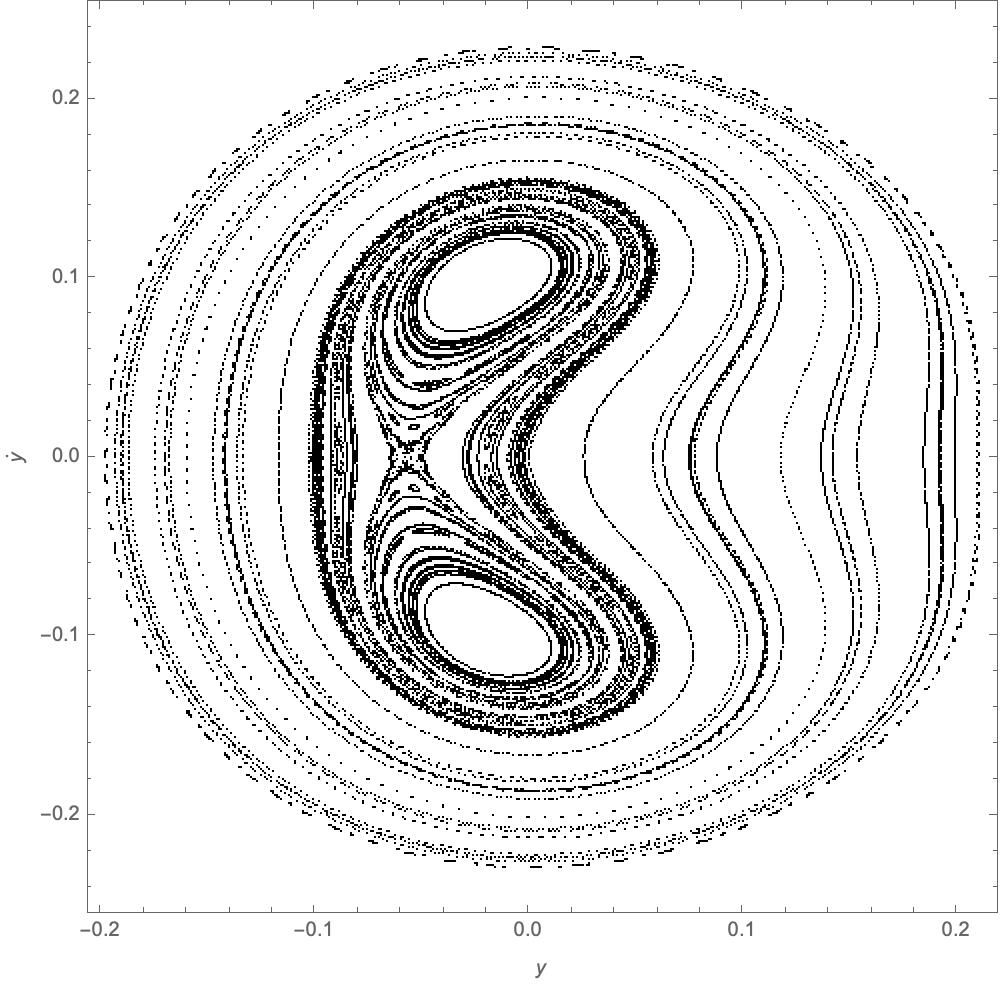}
    }
    \subfloat[   ]{
        \includegraphics[width=0.3\textwidth]{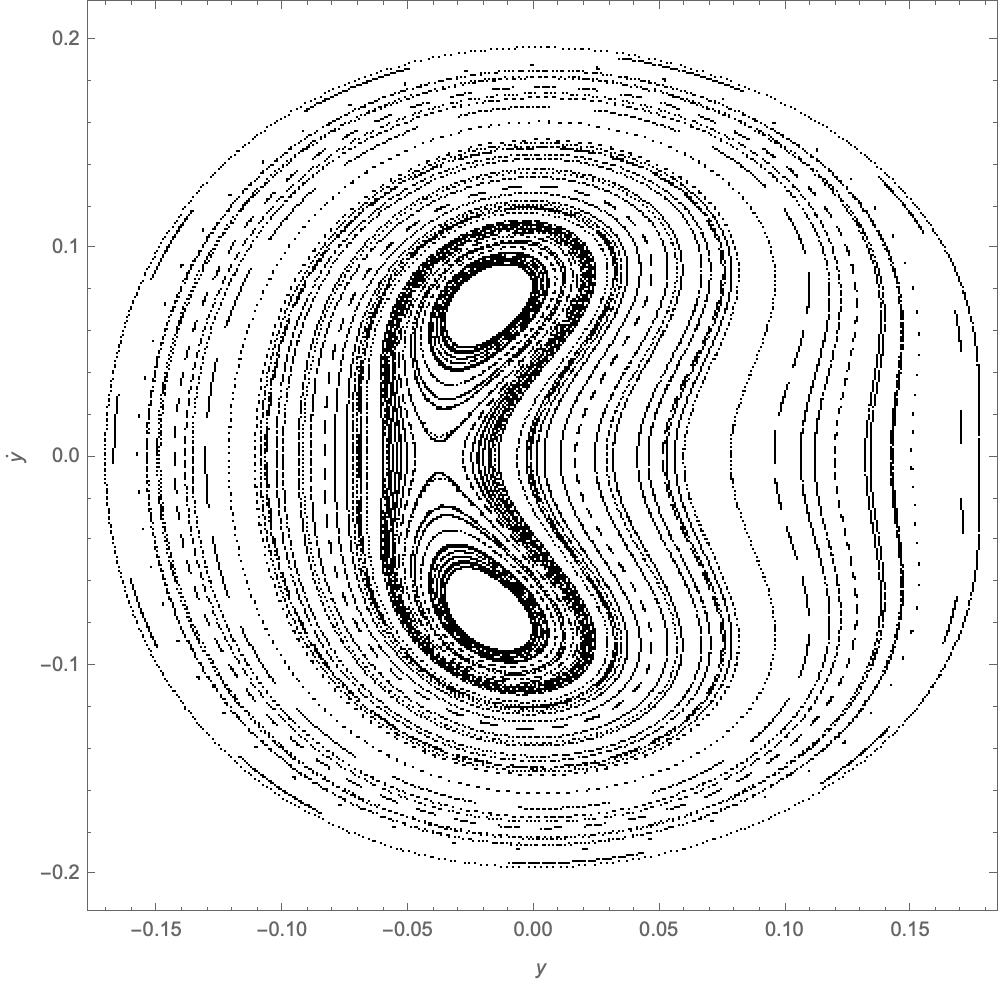}
    }
    \caption{Poincaré sections when $\alpha = 1.0$ and (a) $\delta = 0$ and $n = 1$, (b) $\delta = 0$ and $n = 21$, (c) $\delta = 0$ and $n = 41$, (d) $\delta = 0.1$ and $n = 1$, (e) $\delta = 0.1$ and $n = 21$, (f) $\delta = 0.1$ and $n = 41$, (g) $\delta = 0.5$ and $n = 1$, (h) $\delta = 0.5$ and $n = 21$, (i) $\delta = 0.5$ and $n = 41$, (j) $\delta = 1.0$ and $n = 1$, (k) $\delta = 1.0$ and $n = 21$, (l) $\delta = 1.0$ and $n = 41$ where the energy for each panel is $E = E_{\text{min}}(1-n/100)$.}
    \label{fig:alpha=1.0}
\end{figure}

Next, we find the equations of motion by solving Hamilton's Equation:
\begin{align*}
    \dot{x} &= p_x, \\
    \dot{y} &= p_y, \\
    \dot{p_x} &= -\frac{\partial \mathcal{H}}{\partial x}, \\
    \dot{p_y} &= -\frac{\partial \mathcal{H}}{\partial y}
\end{align*}
where $p_x$ and $p_y$ are the canonical conjugate momenta. After solving these equations, we can start creating our Poincaré sections. For each value of $\alpha$ and $\delta$, we will look at $3$ different energy values so we will end up with $48$ graphs. We split these into four $4 \times 3$ tables of graphs and we get Figures \ref{fig:alpha=0} to \ref{fig:alpha=1.0}.

First, notice Figure \ref{fig:n=1,d=0} which is the Poincaré section of the original HH system. We can see that from the lack of structure and randomness in the plot, the system is chaotic which matches with the findings of Hénon and Heiles. From here, we can change 3 things: $\alpha$, $\delta$, and $E$. For now let $\alpha$ be constant so we first focus on Figure \ref{fig:alpha=0}.

As we go from left to right, we see that more structure seems to emerge. For example, comparing Figure~\ref{fig:n=1,d=0} to Figure \ref{fig:n=41,d=0}, we can see that there are more loops rather than just noise. What this means is that as energy decreases and moves away from $E_{\text{min}}$, the trajectories are less chaotic. Similarly as we go down the table and $\delta$ increases, the chaos decreases. For $\alpha$, we need to compare between different tables. We can use the (c) panel of all tables to see that in Figure \ref{fig:alpha=0}, the Poincaré becomes more chaotic as we go from Figure \ref{fig:alpha=0} to Figure \ref{fig:alpha=1.0}. Therefore the chaos increases as $\alpha$ increases. We can now formulate all these observations into the following points:

\begin{enumerate}
    \item As $E \to E_{\text{min}}$, the chaos increases
    \item As $\delta \to 1.0$, the chaos decreases 
    \item As $\alpha \to 1.0$, the chaos increases
\end{enumerate}

\begin{figure}[h]
    \centering
    \subfloat[]{
        \includegraphics[width=0.22\textwidth]{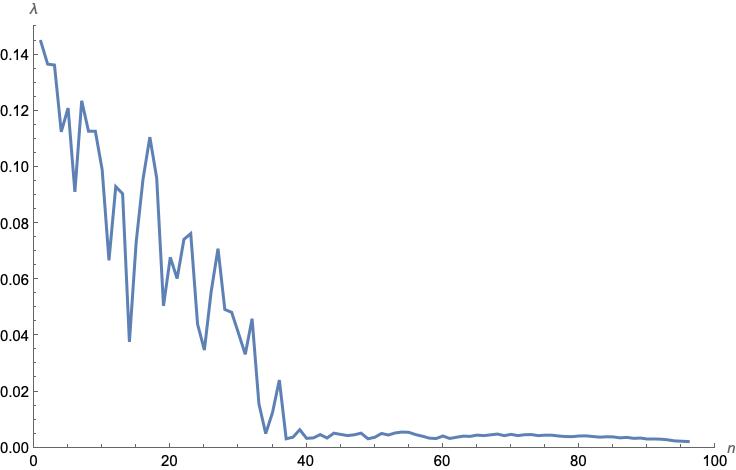}
    }
    \subfloat[]{
        \includegraphics[width=0.22\textwidth]{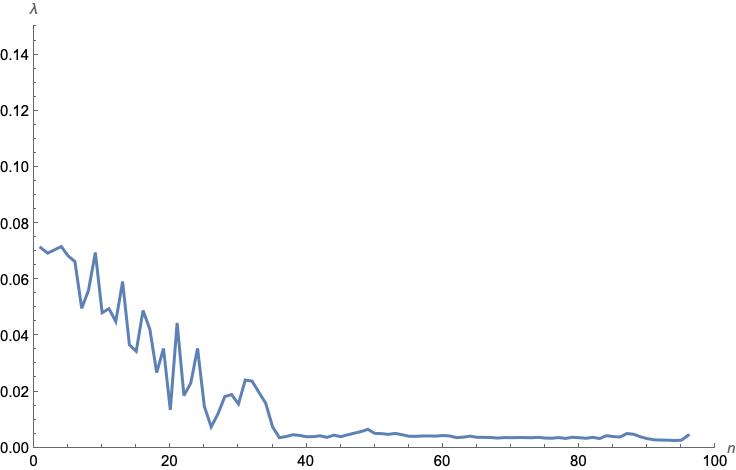}
    }
    \subfloat[]{
        \includegraphics[width=0.22\textwidth]{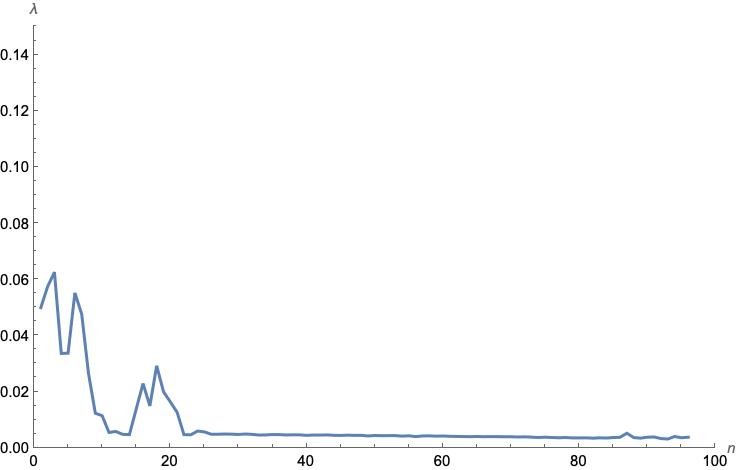}
    }
    \subfloat[]{
        \includegraphics[width=0.22\textwidth]{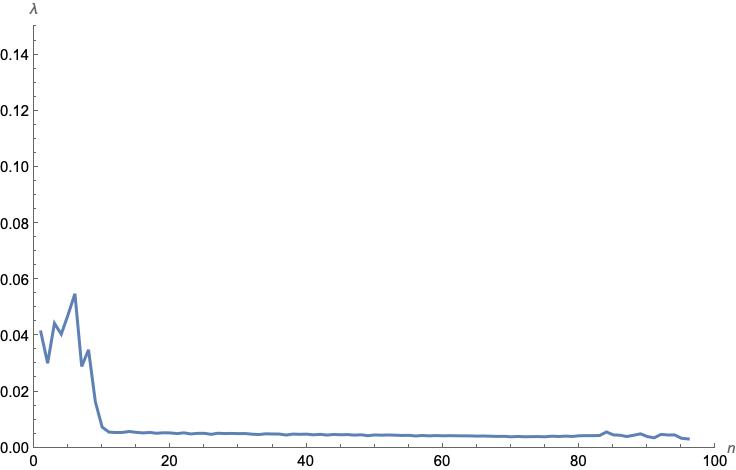}
    }
    \hfill
    \subfloat[]{
        \includegraphics[width=0.22\textwidth]{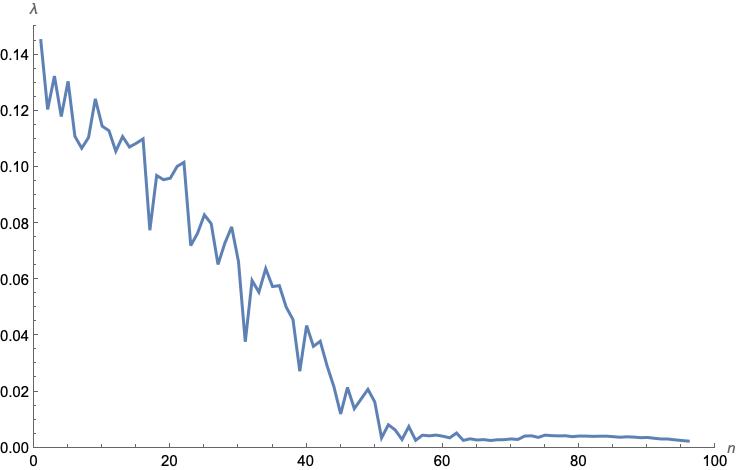}
    }
    \subfloat[]{
        \includegraphics[width=0.22\textwidth]{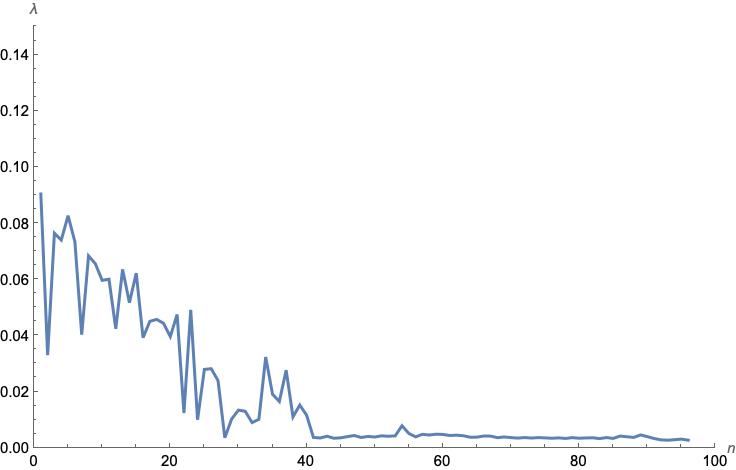}
    }
    \subfloat[]{
        \includegraphics[width=0.22\textwidth]{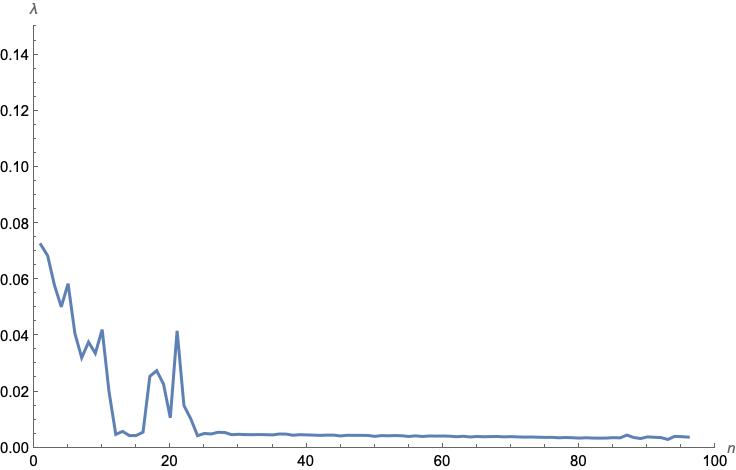}
    }
    \subfloat[]{
        \includegraphics[width=0.22\textwidth]{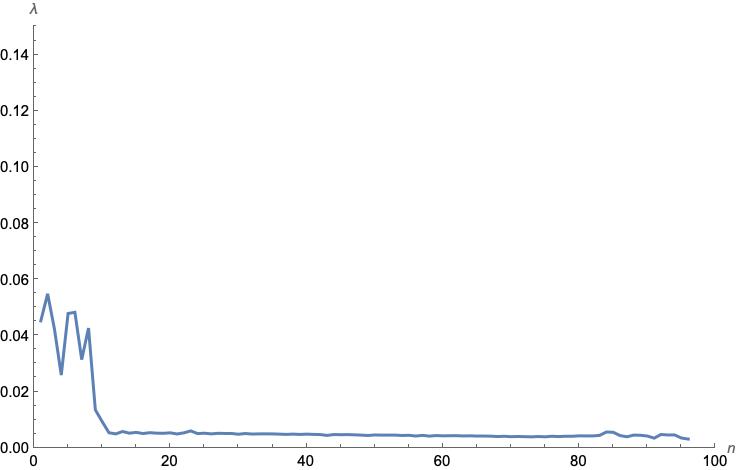}
    }
    \hfill
    \subfloat[]{
        \includegraphics[width=0.22\textwidth]{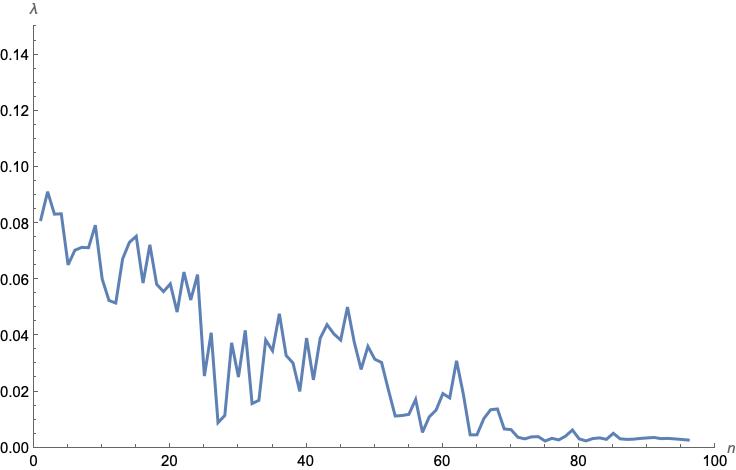}
    }
    \subfloat[]{
        \includegraphics[width=0.22\textwidth]{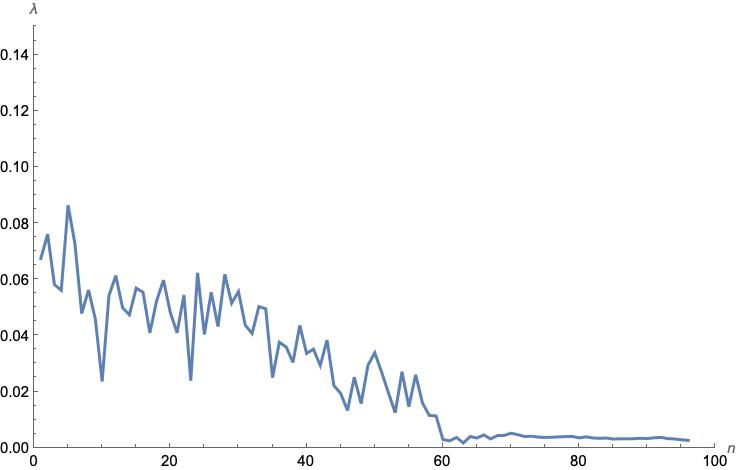}
    }
    \subfloat[]{
        \includegraphics[width=0.22\textwidth]{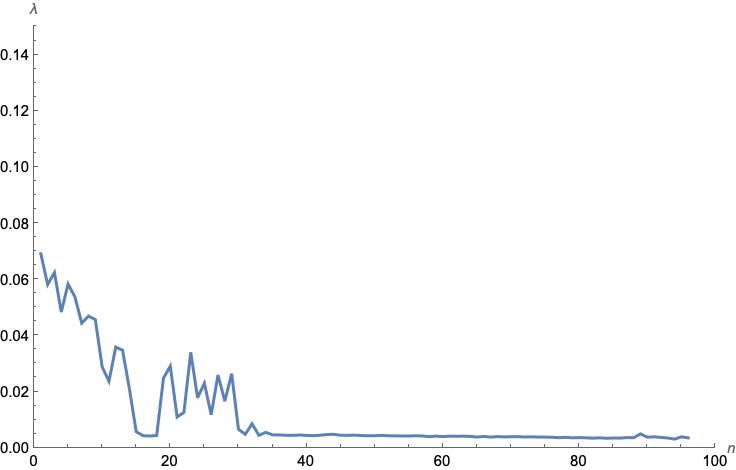}
    }
    \subfloat[]{
        \includegraphics[width=0.22\textwidth]{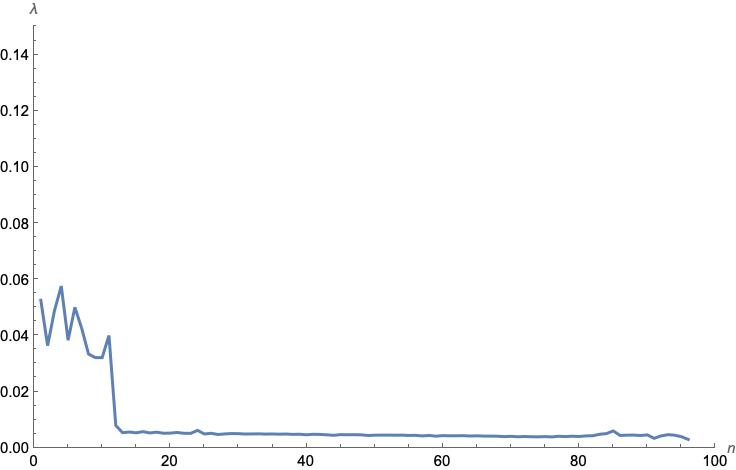}
    }
    \hfill
    \subfloat[]{
        \includegraphics[width=0.22\textwidth]{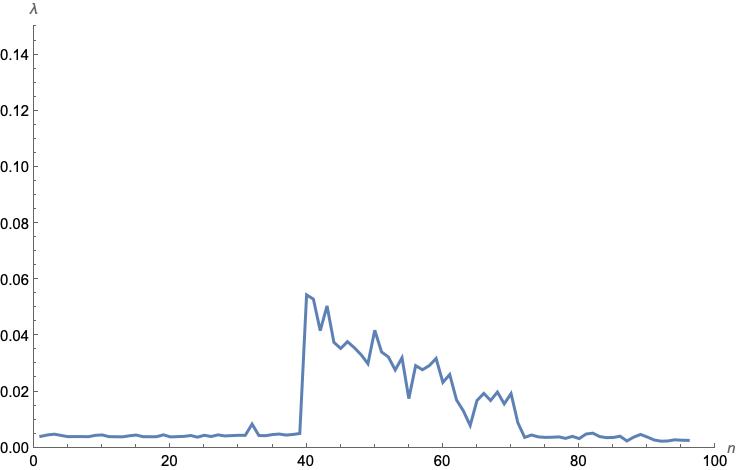}
    }
    \subfloat[]{
        \includegraphics[width=0.22\textwidth]{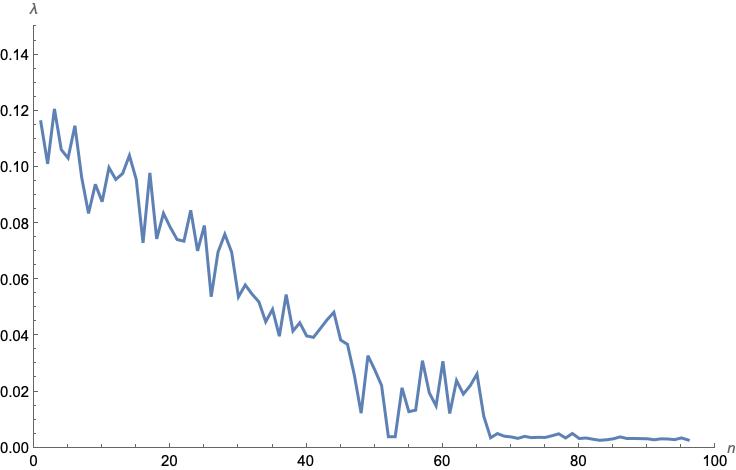}
    }
    \subfloat[]{
        \includegraphics[width=0.22\textwidth]{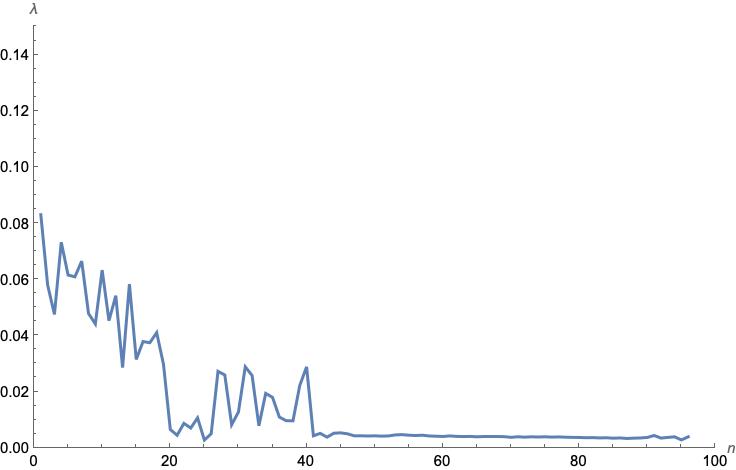}
    }
    \subfloat[]{
        \includegraphics[width=0.22\textwidth]{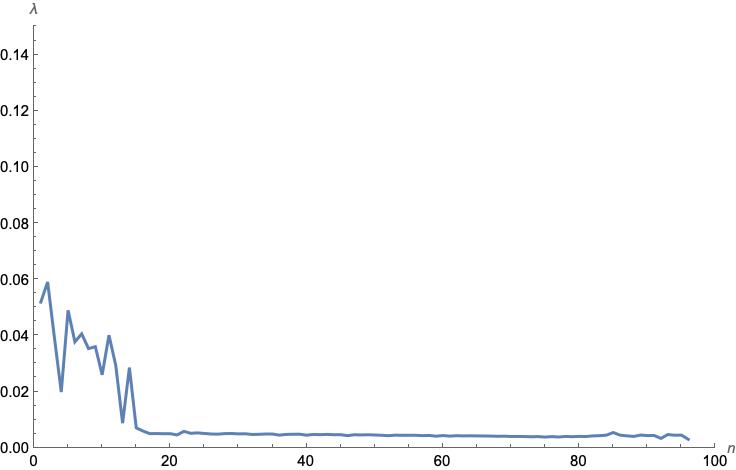}
    }
    \caption{Maximal LCE vs Energy where $\alpha$ is 0, 0.1, 0.5, 1.0 from top to bottom and $\delta$ is 0, 0.1, 0.5, 1.0 from left to right.}
    \label{fig:Lyapunov}
\end{figure}

We can verify these points with the maximal LCE. We plot the maximal LCE, $\lambda$, against the energy, which we parameterize as $E = E_{\text{min}}(1-n/100)$, and we set $x_0 = 0.1$, $y_0=-0.1$, and $\dot{y}_0 = 0.0$ as our initial conditions. See Figure \ref{fig:Lyapunov}.

\begin{figure}[h]
    \centering
    \includegraphics[width=0.45\textwidth]{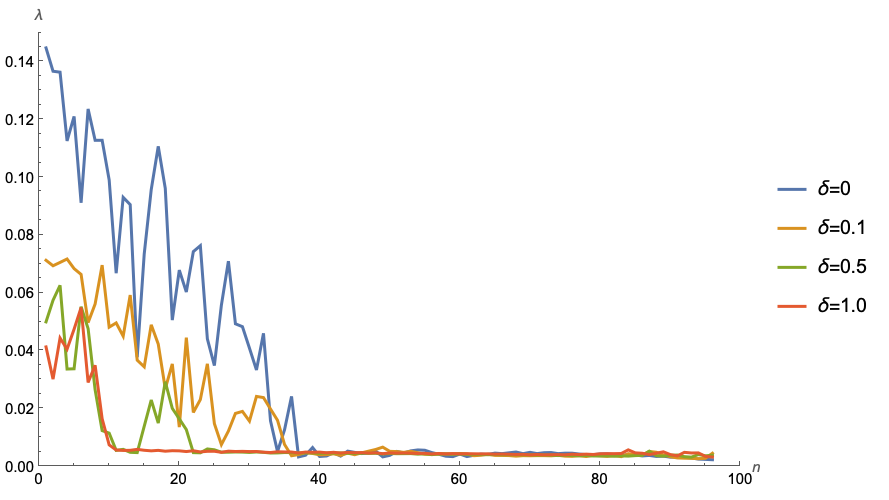}
    \caption{Maximal LCE vs Energy for $\alpha = 0$ and various values for $\delta$}
    \label{fig:delta changing}
\end{figure}

In any of the plots, we can see that as $n \to 0$, equivalent to $E \to E_{\text{min}}$, then $\lambda(n)$, the chaos, increases proving (1) in Theorem 1. Additionally, we can see how $\lambda$ depends on $\delta$ by looking at how $\lambda$ changes at a specific value of $n$ and $\alpha$ when we change $\delta$, say at $n=0$, $\alpha = 0$. See Figure \ref{fig:delta changing}.

\begin{figure}
    \centering
    \includegraphics[width=0.45\textwidth]{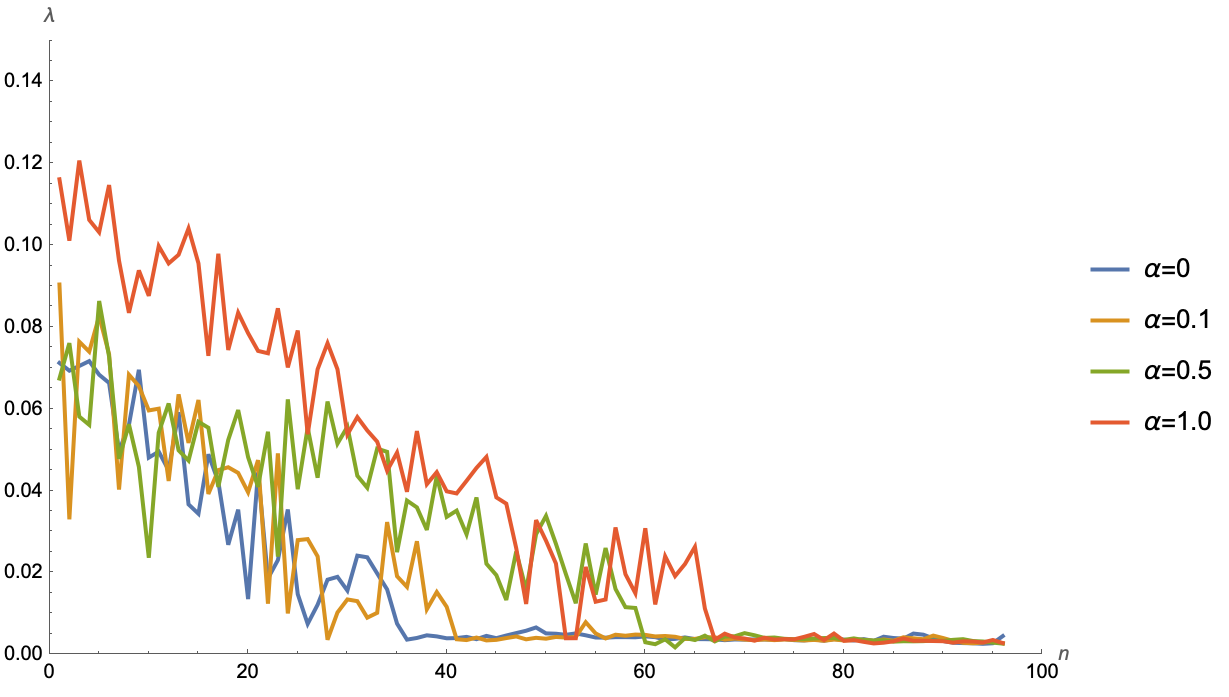}
    \caption{Maximal LCE vs Energy for $\delta = 0.1$ and various values for $\alpha$}
    \label{fig:enter-label}
\end{figure}

We can see that as $\delta \to 1.0$, the height of the graph decreases and so does the Lyapunov exponent. This confirms the second point in the theorem. We can also see this trend in the Poincaré sections. For each 4 by 3 table of Poincaré sections, as we go down the table (which corresponds $\delta \to 1.0$), there is less noise and more structure. We can go through the same reasoning as $\alpha \to 1.0$. As we go through each table, the Poincaré sections develop more noise indicating that the chaos increases. The Lyapunov Exponent also confirms this. See Figure \ref{fig:enter-label}.

\section{Acknowledgments}

Here, the author would like to thank Dr. Fredy L. Dubeibe for inspiring this project, helping him understand the background necessary, and giving feedback for the paper. The author also thanks Garett Brown for helping him understand the numerical aspect of this project. Finally, the author thanks his family for their strong encouragement and support throughout writing this paper.

\bibliographystyle{plain}

\end{document}